\documentclass[useAMS,usenatbib]{mn2e}
\usepackage{amssymb}
\usepackage{graphicx}
\usepackage{xspace}

\usepackage{ifthen}
\def\draftversion{0} 

\makeatletter
\newcommand\mytoc{%
    \@starttoc{toc}%
}
\makeatother

\ifthenelse{\equal{\draftversion}{0}}{
	\usepackage{xcolor}
	\newcommand{\tmp}{}
	\newenvironment{envcomm}[1]{\renewcommand{\tmp}{#1}\begin{color}{blue}\begin{center}\hrule\vspace{0.5mm}\tmp's COMMENTS\end{center}}{\begin{center}END OF \tmp's COMMENTS\vspace{0.5mm}\hrule\end{center}\end{color}}
	\newenvironment{draft}{\begin{color}[rgb]{0,0.4,0}\begin{center}\hrule\vspace{0.5mm}DRAFT\end{center}}{\begin{center}END OF DRAFT\vspace{0.5mm}\hrule\end{center}\end{color}}
	\newcommand{\comcomm}[2]{\begin{color}{blue}\ $\bullet$ \textbf{#1:} #2 $\bullet$\ \end{color}}
	\newcommand{\revend}[1]{\par\begin{color}[rgb]{0,0.4,0}\begin{center}\hrule\vspace{0.5mm}END OF #1's REVISIONS\vspace{0.5mm}\hrule\end{center}\end{color}\par}
	\newcommand{\todo}[1]{\begin{color}{red}\ $\bullet$ \textbf{To do: }#1 $\bullet$\ \end{color}}
	
	\newcommand{\del}[1]{\begin{color}[rgb]{0,0.5,0.0}\ $\bullet$ \textbf{Deleted: }#1 $\bullet$\ \end{color}}
	\newcommand{\sk}[1]{\begin{color}[rgb]{0.6,0,0.6}#1\end{color}}
	\newcommand{\toc}{\par\begin{color}[rgb]{0.6,0,0.6}\begin{center}\hrule\vspace{0.5mm}\begingroup\small\let\cleardoublepage\relax\let\clearpage\relax\mytoc\endgroup\vspace{0.5mm}\hrule\end{center}\end{color}\par}
	}{
	\newsavebox{\trashcan}
	\newenvironment{envcomm}[1]{\begin{lrbox}{\trashcan}\begin{minipage}{\columnwidth}}{\end{minipage}\end{lrbox}}
	
	\newcommand{\comcomm}[2]{}
	\newcommand{\revend}[1]{}
	\newcommand{\todo}[1]{}
	
	\newcommand{\del}[1]{}
	\newcommand{\sk}[1]{}
	\newcommand{\toc}{}
	}

\newcommand{\aj}{AJ}
\newcommand{\araa}{ARA\&A}
\newcommand{\apj}{ApJ}
\newcommand{\apjl}{ApJ}
\newcommand{\apjs}{ApJS}
\newcommand{\aap}{A\&A}
\newcommand{\mnras}{MNRAS}
\newcommand{\na}{New A}
\newcommand{\pasj}{PASJ}

\newcommand{\mh}{\ensuremath{\textrm{\,--\,}}}
\newcommand{\bb}[1]{\ifmmode \mbox{\boldmath $ #1$} \else  \mbox{\boldmath $#1$} \fi}

\newcommand{\U}[1]{\ensuremath{\mathrm{~#1}}}

\newcommand{\yr}{\U{yr}}
\newcommand{\Myr}{\U{Myr}}
\newcommand{\Gyr}{\U{Gyr}}
\newcommand{\pc}{\U{pc}}
\newcommand{\kpc}{\U{kpc}}
\newcommand{\Mpc}{\U{Mpc}}
\newcommand{\msun}{\U{M}_{\odot}}
\newcommand{\Msun}{\msun}
\newcommand{\Msunyr}{\Msun\yr^{-1}}
\newcommand{\cc}{\U{cm^{-3}}}
\newcommand{\kms}{\U{km\ s^{-1}}}

\newcommand{\hii}{H{\sc ii} }

\newcommand{\rh}{\ensuremath{r_\mathrm{h}}}

\newcommand{\eqn}[2][]{Equation#1~\ref{eqn:#2}} 
\newcommand{\fig}[2][]{Figure#1~\ref{fig:#2}}
\newcommand{\tab}[2][]{Table#1~\ref{tab:#2}}
\newcommand{\sect}[2][]{Section#1~\ref{sec:#2}}
\renewcommand{\eqn}[2][]{equation#1~(\ref{eqn:#2})}
\renewcommand{\fig}[2][]{Fig#1.~\ref{fig:#2}}

\newcommand{\citetip}[1]{#1 (in prep.)}
\newcommand{\citepip}[1]{(#1, in prep.)}
\newcommand{\citealtip}[1]{#1, in prep.}

\title[A parsec-resolution simulation of the Antennae galaxies]{A parsec-resolution simulation of the Antennae galaxies:\\Formation of star clusters during the merger}

\author[Renaud, Bournaud \& Duc]{Florent~Renaud\thanks{florent.renaud@cea.fr}, Fr\'ed\'eric~Bournaud and Pierre-Alain~Duc\\
Laboratoire AIM Paris-Saclay, CEA/IRFU/SAp, Universit\'e Paris Diderot, F-91191 Gif-sur-Yvette Cedex, France
}

\begin{document}
\maketitle

\newcommand{\Bournaud}{Bournaud et al.}
\newcommand{\Duc}{Duc et al.}
\newcommand{\Gieles}{Gieles \& Renaud}
\newcommand{\Emsellem}{Emsellem et al.}

\begin{abstract}
We present a hydrodynamical simulation of an Antennae-like galaxy merger at parsec resolution, including a multi-component model for stellar feedback and reaching numerical convergence in the global star formation rate for the first time. We analyse the properties of the dense stellar objects formed during the different stages of the interaction. Each galactic encounter triggers a starburst activity, but the varying physical conditions change the triggering mechanism of each starburst. During the first two pericenter passages, the starburst is spatially extended and forms many star clusters. However, the starburst associated to the third, final passage is more centrally concentrated: stars form almost exclusively in the galactic nucleus and no new star cluster is formed. The maximum mass of stars clusters in this merger is more than 30 times higher than those in a simulation of an isolated Milky Way-like galaxy. Antennae-like mergers are therefore a formation channel of young massive clusters possibly leading to globular clusters. Monitoring the evolution of a few clusters reveals the diversity of formation scenarios including the gathering and merger of gas clumps, the monolithic formation and the hierarchical formation in sub-structures inside a single cloud. Two stellar objects formed in the simulation yield the same properties as ultra-compact dwarf galaxies. They share the same formation scenario than the most massive clusters, but have a larger radius either since birth, or get it after a violent interaction with the galactic center. The diversity of environments across space and time in a galaxy merger can account for the diversity of the stellar objects formed, both in terms of mass and size.
\end{abstract}
\begin{keywords}galaxies: interactions --- galaxies: starburst --- galaxies: star clusters: general --- galaxies: star formation --- ISM: structure --- methods: numerical\end{keywords}

\section{Introduction}

Interacting galaxies are known to trigger bursts of star formation, although this statement is toned down when considering selection biases and the weak contribution of mergers relative to main sequence galaxies to star formation in the Universe \citep{Wolf2005}. Nevertheless, a merger event remains a major milestone in the star formation history of a given galaxy. Since mergers provide a wide range of physical conditions for star formation, it is likely that they also account for the diversity of observed dense stellar systems, from associations to globular clusters and possibly ultra compact dwarf galaxies. What type of stellar objects can be formed in a merger, as opposed to a more regular formation in an isolated galaxy, and under which conditions, remains poorly understood.

For several decades, simulations have been an efficient tool to study the morphology of interacting galaxies and mergers, and comparing the physics of their star formation with isolated cases. Studies first determined the tidal origin of the tails and the influence of orbital parameters on the morphology and dynamics of the collision \citep[e.g.][]{Toomre1972, Barnes1988, Naab2003}. Hydrodynamical simulations established a link between pericenter passages and bursts of star formation (lasting a few $\times 10 \Myr$) and have began to explore the properties of the interstellar medium (ISM) and the formation of massive star clusters in mergers \citep{Mihos1993, Barnes2004, DiMatteo2007, Cox2008, Bournaud2008, Saitoh2009, Karl2010, Teyssier2010, Powell2013}. 

Because of technical limitations, previous simulations were not able to fully probe the scale of star forming clouds ($\lesssim 10 \pc$). However, the coupling between the kpc-scale dynamics and the physics of star formation has been established using high resolution simulations of isolated galaxies \citep{Bournaud2010b, Renaud2013b, Fujimoto2014}. Using simulations of mergers, \citet{Teyssier2010} underlined the dependance of the simulated star formation rate (SFR) with resolution, and thus the importance of resolving parsec-scale star forming structures. However, the computational cost of considering two galaxies, their tidal features spanning huge volumes and the highly fragmented nature of the ISM prevented them to reach numerical convergence in their estimate of the SFR\footnote{Numerical convergence is reached when the estimate of a physical quantity measured in a simulation becomes independent of numerical resolution.}. (Their simulation reached a resolution of $12 \pc$, and did not include stellar feedback.)

Other numerical works focussed on the formation of star clusters in mergers, highlighting the possible formation of very massive clusters in the nuclear regions of remnants or during the collision stages \citep{Bekki2001, Matsui2012}, emphasising relations between the radial position of clusters and their internal properties like metallicity \citep{Bekki2002} and age \citep{Li2004}, and proposing mergers to be one formation channel of massive globulars \citep{Bournaud2008}. These pioneer simulations faced the technical challenges of resolving star clusters and probing their intermediate mass regime ($\sim 10^{4\mh 5} \Msun$). With the present day computational ressources, we are able to push these limits to the parsec scale, smaller masses and account for more detailed physics. 

The Antennae galaxies (NGC~4038/39, Arp~244) have received a particular attention, both from the observational and the numerical sides, being the closest major merger in the starburst phase ($\sim 20 \Mpc$, \citealt{Whitmore1999a}, see also \citealt{Saviane2008, Schweizer2008}). Several regions, and not only the galactic nuclei, host the formation of very massive stellar objects ($\gtrsim 10^6 \Msun$, $\sim 10 \Myr$, \citealt{Whitmore1995, Mengel2005, Whitmore2007, Bastian2009, Whitmore2010, Herrera2011}). In total, more than 1000 star clusters have been detected in the central $\sim 10 \kpc$ of the Antennae. The richness of this system and the profusion of observational constraints make the Antennae an excellent benchmark for theories and simulations on the role of merger in (re-)shaping the structural properties of galaxies and of their stellar populations.

In this paper, we present a hydrodynamical simulation of the Antennae galaxies, at $1.5 \pc$ resolution, including star formation and stellar feedback (photo-ionisation, radiative pressure and supernovae). The gain from previous works in term of resolution and physics implemented allows us to better describe the formation and early evolution of stellar systems by probing smaller objects and resolving the inner structure of the most massive cases. \citet{Renaud2014a} used this simulation to establish a relation between the compressive nature of the tides and the interstellar turbulence in mergers and the enhancement of star formation. Here, we focus our analysis on the properties of the stellar populations formed during the several stages of the interaction. We first present the numerical techniques adopted (\sect{numerics}) and the model of the Antennae (\sect{model}). \sect{classification} details the classification of the stellar objects detected in the simulation. Their formation and early evolution is presented in \sect{formation}. The cluster mass function is discussed in \sect{cmf}. \sect{variety} compares the properties of the stellar objects with observations and the formation history of a few objects is followed across time to provide possible scenarios for the formation of equivalent systems in the reality. Many aspects of this simulation are being analysed to address questions about the nature of the interstellar medium and to compare the properties of the modelled systems to the real Antennae galaxies adopting an observational perspective. These will be presented in forthcoming contributions (\citealtip{\Bournaud}, \citealtip{\Duc}).

\section{Numerical technique}
\label{sec:numerics}

The simulation has been run using the adaptive mesh refinement (AMR) code RAMSES \citep{Teyssier2002}. The overall method and parameters are comparable to those used in the simulation of the Milky Way by \citet{Renaud2013b}. 
The main difference is the use of a thermodynamical model including heating by ultraviolet radiation of cosmic origin, and atomic cooling tabulated at solar metallicity by \citet{Courty2004}, as in \citet{Perret2014}. The minimum allowed temperature is set to 50 K to prevent numerical artefacts \citep{Truelove1997, Robertson2008}. The use of such heating/cooling scheme allows for deviations from the otherwise simpler polytropic equation of state of \citet{Renaud2013b}, in particular at low densities and high temperatures, a common situation in galaxy mergers where tidal interactions can eject a significant fraction of the ISM from the galactic discs.

Gas denser than $50 \cc$ is converted into stellar particles (called stars hereafter, for simplicity) of minimum mass of $70 \Msun$, with an efficiency of 2 per cent per free-fall time. This sets the star formation rate to $\sim 1 \Msunyr$ before the interaction. The newly formed stars inject feedback energy in the form of photo-ionisation in \hii regions, radiation pressure \citep{Renaud2013b} and type-II supernova thermal blasts \citep{Dubois2008, Teyssier2013}. To speed-up the computation, only one out of ten young stellar particles ionises its surrounding ISM, but with an energy ten times higher than a single source. Since stars form in a clustered way in the dense gas regions, each of these regions hosts at least one ionising source, such that our simplification does not alter the global energy budget at the scale of molecular clouds.

The refinement strategy adopted is ``quasi-Lagrangian'' (as in \citealt{Teyssier2010, Renaud2013b, Bournaud2014}): an AMR cell is refined when it contains more than a given number ($\sim 50$) of particles (dark matter and stars from the initial conditions), or when its baryonic density (stars and gas) exceeds a certain value (which obviously depends on the refinement level of the cell). Furthermore, the code ensures that the Jeans length is always resolved by at least four cells. The maximum resolution is $1.5 \pc$ in the most refined volumes of the $(200 \kpc)^3$ computational domain. The dark matter and stellar components from the initial setup (as opposed to the stars formed during the simulation) of each of the two galaxies are rendered with $\sim 2.3$ millions particles and evolved using a particle-mesh technique at the resolution of $50 \pc$. These components provide a ``live'' background potential on top of which the ISM and the young stellar objects can evolve. The stars formed from the gas are evolved using the resolution of the AMR grid, i.e. up to $1.5 \pc$.

Due to the softening of the gravitational force on the stars, the two-body interaction is not properly described in our simulation and related processes could not be probed. For instance the internal evolution of star clusters driven by two-body relaxation is not accounted for in this work. We argue later that the relaxation time of such objects is much longer than the time-lapse considered in our analysis, i.e. mainly the duration of the galactic interaction (see \sect{cfr}).

The simulation has been run for eight million core hours on 4096 cores of the supercomputer SuperMUC hosted at the \emph{Leibniz-Rechenzentrum}.

\section{Model of the Antennae galaxies}
\label{sec:model}

\subsection{Initial conditions and orbit}
\label{sec:orbit}

\begin{table}
\caption{Initial setup}
\label{tab:init}
\begin{tabular}{lcc}
\hline
& NGC~4038 & NGC~4039 \\
\hline
\multicolumn{3}{l}{Orbit (on the plane of the sky)}\\
x position [kpc] & -15.0 & 10.6 \\
y position [kpc] & 30.4 & -30.3\\
z position [kpc] & -46.3 & 46.7\\
x velocity [km/s] & 26.0 & -26.9 \\
y velocity [km/s] & -23.3 & 23.2\\
z velocity [km/s] & 71.3 & -71.8\\
x spin axis & 0.65 & 0.67 \\
y spin axis & 0.65 & -0.71\\
z spin axis & -0.40 & 0.20\\
total mass [$\times 10^{11} \msun$] & 3.10 & 3.10\\
\hline
\multicolumn{3}{l}{Gas disc (exponential)}\\
mass [$\times 10^9 \msun$] & 6.5 & 5.0\\
characteristic radius [kpc] & 6.5 & 3.9\\
truncation radius [kpc] & 16.0 & 10.5\\
characteristic height [kpc] & 0.4 & 0.4\\
truncation height [kpc] & 1.5 & 1.2\\
\hline
\multicolumn{3}{l}{Stellar disc (exponential)}\\
number of particles & 1,006,000 & 1,006,000\\
mass [$\times 10^9 \msun$] & 47.8 & 47.8\\
characteristic radius [kpc] & 3.5 & 4.0\\
truncation radius [kpc] & 11.0 & 12.0\\
characteristic height [kpc] & 0.35 & 0.35\\
truncation height [kpc] & 1.0 & 1.0\\
\hline
\multicolumn{3}{l}{Stellar bulge (Hernquist)}\\
number of particles & 372,000 & 372,000\\
mass [$\times 10^9 \msun$] & 17.6 & 17.6\\
core radius [kpc] & 0.8 & 0.8\\
truncation radius [kpc] & 2.5 & 2.5\\
\hline
\multicolumn{3}{l}{Dark matter halo (Burkert)}\\
number of particles & 1,000,000 & 1,000,000\\
mass [$\times 10^9 \msun$] & 237.6 & 237.6\\
core radius [kpc] & 10.0 & 10.0\\
truncation radius [kpc] & 35.0 & 35.0\\
\hline
\end{tabular}
\end{table}

The present simulation has been tailored to reproduce the morphology and kinematics of the Antennae galaxies (NGC~4038/39, Arp 244). It is based on the initial parameters from \citet{Renaud2008}, found with a try-and-see approach where the quality of the model is estimated by eye, and not relying on numerical techniques to optimise the match with observational data (for such studies, see e.g. \citealt{Privon2013}, \citealt{Karl2013}). This model can be used to analyse the physical processes of similar prograde-prograde major mergers, as opposed to other configurations producing different features and hosting different populations (see e.g. \citealt{Bournaud2008}).

For simplicity, we adopt the naming of the real Antennae when referring to our simulated galaxies: NGC~4038 is the Northernmost galaxy at present time, associated with the Southern tidal tail. Since we observe the Antennae between their second and third pericenter passages, and that the two galaxies switch from North to South at each passage, NGC~4038 is also North initially (i.e. before the first collision). The initial parameters for the progenitors and their orbit are given in Table~\ref{tab:init}. The two progenitor galaxies have been initialized as disc galaxies, with almost exactly the same mass. The dark matter halo follows a \citet{Burkert1995} profile truncated at the radius when its reaches five times the stellar disc mass. The total baryonic fraction is eight per cent. The stellar component is made of a \citet{Hernquist1990} bulge which represents 27 per cent of the stellar mass, and an exponential disc. The initial gas fraction is $\sim 9$ per cent.

We ensure that no non-axisymetric structure like a bar or spirals is formed in the discs before the first collision. Such structures may create a phase reference point which would play an important role in the shaping of the tidal features, and the transfer of angular momentum within the central regions of the discs. We note however that the collision itself accelerates the formation of a bar in both our discs, as often seen in simulations of interacting galaxies \citep[e.g.][]{Berentzen2004, Cox2008, Hopkins2009}.

The orbit is comparable to that of \citet{Renaud2008} and \citet{Teyssier2010}, although the progenitors are initially separated from a larger distance than in these previous works. Our simulation is stopped at the beginning of the final coalescence of the galaxies. Following the evolution for a longer time would be numerically costly and may require to account more precisely for cosmological aspects like the environment of the pair \citep{Moreno2013}.

\subsection{A Milky Way reference model}
\label{sec:mw}

In the following, stellar objects in the galaxy merger are compared to those detected, using the same methods, in the simulation of the Milky Way (MW) presented in \citet{Renaud2013b}. The maximum resolution of this simulation is $0.05 \pc$, i.e. 32 times higher than that of the Antennae. However, only a few Myr have been simulated at this resolution. To ease the comparison with the Antennae, we use a slightly earlier snapshot from the MW, when the resolution is set to $1.5 \pc$ (i.e. $5 \Myr$ before the instant pictured in the Figure 4 of \citealt{Renaud2013b}), but already when the large-scale structures (spirals and bar) of the Galaxy are in place and host star formation.

The two simulations differ in term of the cooling scheme adopted: the Milky Way simulation uses a piecewise polytropic equation of state while ultraviolet heating and atomic cooling are invoked for the Antennae run. \citet{Kraljic2014} demonstrated that the difference between the two schemes is negligible, at least in the dense, star-forming gas.

Therefore, the physical properties of the stellar objects from both simulations are directly comparable.

\subsection{Star formation history}
\label{sec:sfr}

\begin{figure*}
\includegraphics{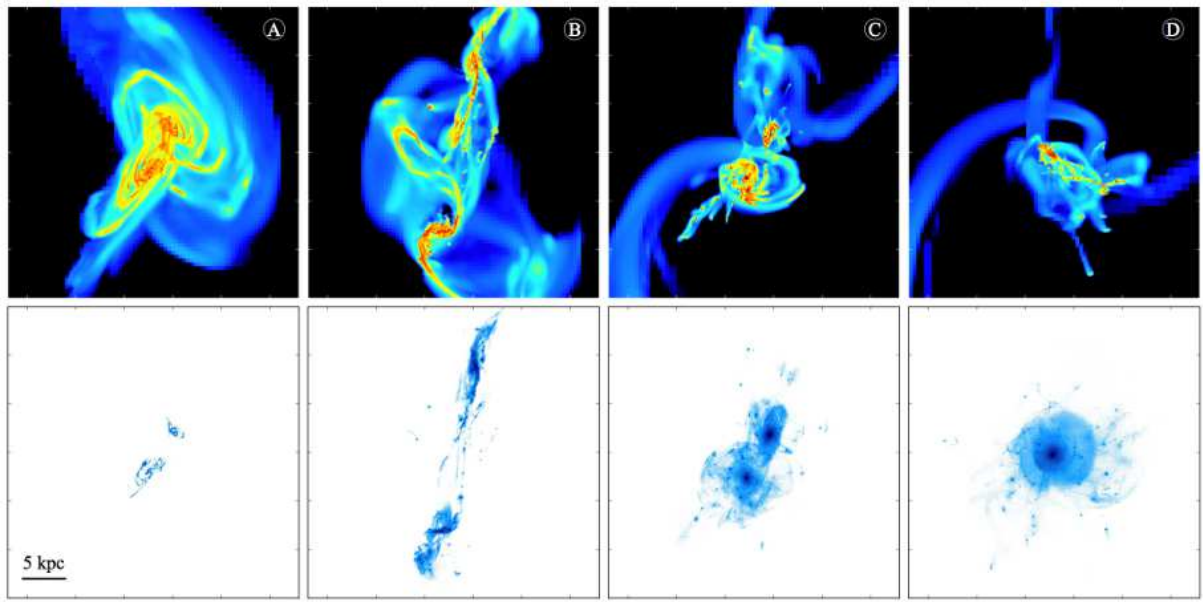}
\caption{Surface density maps of the gas component (top) and the stars formed in the simulation (bottom) at four key stages of the interaction, in the central $30\kpc \times 30 \kpc$ seen in the ``plane of sky''. From left to right, the epochs corresponds to $t = -6 \Myr$, $44 \Myr$, $137 \Myr$ and $192 \Myr$. These maps reveal the off-nuclear formation of star clusters, in particular in the tidal bridge connecting the two galaxies while they separate, and the elongation of stellar objects in streams and shells at coalescence.}
\label{fig:morphology}
\end{figure*}

The reference time $t=0$ corresponds to the first pericenter passage, occurring $432 \Myr$ after the start of the simulation. A long pre-collision phase is required to eliminate possible artefacts in the initial setup of the galactic discs, and for the galaxies to start separated by a sufficient distance so that they would not experience strong tidal effects at setup.

\fig{morphology} shows a sequence of snapshots extracted at various time along the collision.\footnote{Movies of the simulation are available here: http://irfu.cea.fr/Pisp/florent.renaud/movies.php} Contrary to the tidal tails which do not host much star formation, the gaseous component of the tidal bridge (i.e. the elongated structures of tidal origin, connecting the two galaxies) rapidly fragments into dense clumps. Self-gravity takes over the strong extensive (i.e. destructive) tidal field, the clouds collapse and further form stars in a compact, clustered fashion. The clustered nature of star formation, followed by the elongation of these structures by tidal and orbital effects is visible in the outer parts of the progenitors. Some of the gas clouds and star clusters remain close to the gravitational Lagrange point of equilibrium between the two galaxies, but most of them fall back at high speed (up to $300 \kms$) toward one or the other progenitor. Note that this process allows for material from one disc to be accreted onto the other. Although this has little importance in our simulation, it may be a key aspect for mixing the gaseous and stellar material when very different galaxies are involved in the merger (e.g. non-coeval or with different metallicities, SFRs etc.).

\begin{figure*}
\includegraphics{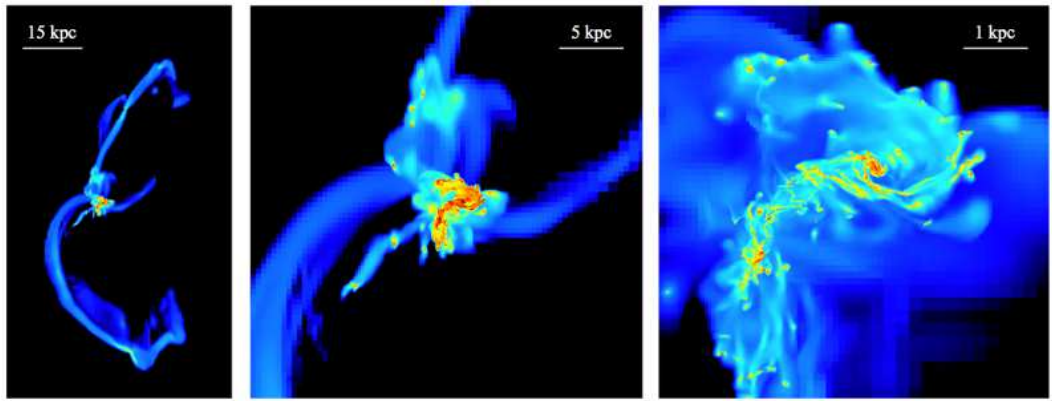}
\caption{Density maps of the gas component as seen in the ``plane of sky'', when the model best matches the observations ($t = 152 \Myr$). The right-hand side view shows the well-known features of the overlap and the Northern arc both containing many dense and massive star forming clouds.}
\label{fig:morpho_gas}
\end{figure*}

\fig{morpho_gas} shows the large scale morphology of the gas component when it best matches the observations of the Antennae. This instant is reached at $t \approx 152\Myr$, i.e. $6 \Myr$ after the second pericenter passage and the moment when the two discs start to overlap \citepip{\Duc}, and a few Myr before the third (and last) passage. The simulation is run until the beginning of the final coalescence. The much higher level of fragmentation of the ISM at this late stage, compared to the pre-interaction epoch, increases significantly the numerical cost of the simulation, such that it becomes very expensive to pursue the integration of the merger remnant.

\begin{figure}
\includegraphics{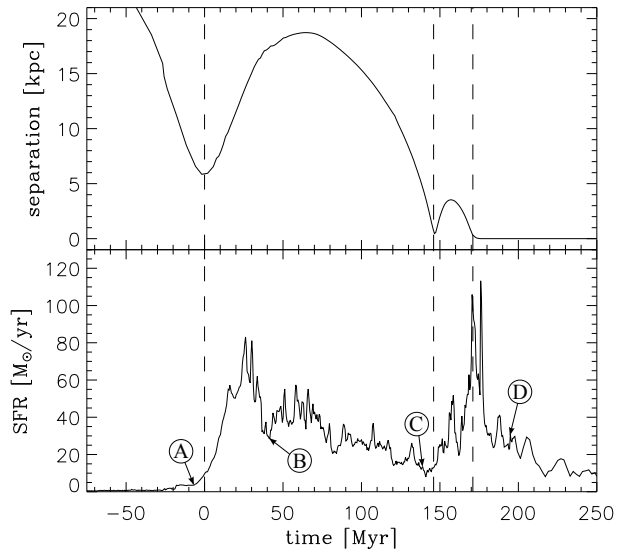}
\caption{Evolution of the separation between the two progenitors (set manually to zero for $t > 175 \Myr$ because of the difficulty to distinguish density centers just before coalescence), and SFR during the interaction ($t=0$ marks the first pericenter passage). Bursts of star formation are associated to the pericenter passages (vertical lines) and the beginning of the final coalescence ($t \approx 180 \Myr$). This curve has been smoothed with a $2 \Myr$ wide Gaussian window, to eliminate the high frequencies, for the sake of visibility. The letters A, B, C and D label several epochs used later in our analysis, as in \fig{morphology}. The best match with the observations is reached at $t = 152 \Myr$, and yields a SFR of $\approx 22 \Msunyr$.}
\label{fig:sfr}
\end{figure}

\fig{sfr} displays the evolution of the SFR along the interaction. As noted by numerous previous studies, episodes of intense star formation occur at the times of pericenter passages and at the beginning of the final coalescence of the two galaxies \citep{Barnes1991, Mihos1996, Cox2006, DiMatteo2007, Karl2010, Teyssier2010}. The details of the bursts (shape, intensity) depend on the physics resolved by the simulation, in particular the ability to capture the turbulence cascade \citep{Teyssier2010}. Using the simulation presented here and the same setup at lower resolutions, we have shown in  \citet{Renaud2014a} that the global SFR becomes independent of the numerical resolution when it reaches $\sim 6 \pc$. We also explained that compressive tides generated by the galactic encounters inject turbulent energy into the ISM. In such turbulence, the compressive component overtakes the solenoidal one, which creates an excess of dense gas, leading to an intense episode of star formation. This process triggers star formation not only in the central regions of galaxies but also over extended volumes, several kpc away from the galactic nuclei.

After the first starburst, the SFR drops by a factor of 2.5 ($t \approx 45 \Myr$) before rising again within $\sim 10 \Myr$, and slowly declining as the progenitors separate. This drop corresponds to the end of the star formation in the tidal debris, in particular in the tidal bridges between the two galaxies. This results from the lowering of the gas density in these regions because of both consumption to form stars, and dynamical evolution of the cloud leading to its elongation and dilution of gas. Large volumes within the discs still host star formation, which maintains a global SFR $\gtrsim 30 \Msunyr$. A few Myr later, tidal debris from the bridges and the tails fall back into the galaxies, either collide with other gaseous structures or simply get compressed along their orbit. This fuels a secondary episode of star formation, at a lower rate than the initial starburst: $\sim 40 \Msunyr$ on average, with short and localized bursts reaching $\sim 60 \Msunyr$. This process continues during the entire separation of the galaxies and the global SFR decreases to $\sim 10 \Msunyr$ just before the second pericenter passage. 

The second encounter also triggers a rapid rise of the SFR, but at this stage, it becomes involved to identify the peaks of the global SFR and connect them to their physical cause. Dynamical friction prevents the galaxies to separate significantly after the second passage: the maximum distance between the nuclei is $3.5 \kpc$, i.e. less than the radii of the stellar discs. In other words, the two gas reservoirs continuously overlap for $t \gtrsim 150 \Myr$. Gravitational torques drive inflows of gas toward the nuclei \citep{Barnes1991, Barnes1996}. At the same time, collisions between gas structures (spiral arms, clumps) occur more frequently than at the first passage because of a smaller impact parameter \citep{Jog1992}. These processes, in addition to the compression by tides and turbulence \citep{Renaud2014a}, trigger a new starburst episode. The peak of the SFR corresponds to the merger of the two nuclei, leading to a short ($\sim 10 \Myr$) but intense (up to $110 \Msunyr$) star formation episode. This point is further discussed in \sect{cfr}.

The SFR curve in \fig{sfr} already indicates the very different nature of the two main bursts. In our setup, both bursts of star formation yield, at first order, an exponential decline with characteristic timescales of $90$ and $30 \Myr$ respectively. Furthermore, the full burst episode, i.e. the rather sharp rise of the SFR followed by a slower decline can be roughly fitted by a log-normal functional form. In that case, the skewness of the function is about three times larger for the first burst than for the second. This shows the differences between (i) the long-range effect of the tidal compression at the first burst, (ii) the slow decrease of the SFR during the separation of the progenitor, and (iii) the direct collision between the discs and their nuclei at the second and third encounters.

After the third passage, the nuclei start to coalesce. The damped modulation in the peaks of the SFR for $t > 170 \Myr$ corresponds to the oscillations of the two nuclei around each other, while dynamical friction makes them loose orbital energy, making each nucleus-nucleus passage less and less efficient at triggering a central starburst. Note that the same behaviour is commonly found in the mass fraction in compressive tides \citep{Renaud2009}. Although the numerical cost of the simulation prevents us from evolving the system for a long post-coalescence period, a complementary run at the resolution of $3 \pc$ indicates that the SFR resumes its pre-interaction level at $t \approx 300 \Myr$, i.e. $120 \Myr$ after the last starburst. Therefore, when including the rise of SFR just before the first passage, the full episode of enhanced star formation lasts $320 \Myr$.

The best match with observations being reached at $t=152 \Myr$, we note that the SFR has declined and increased again before reaching the present day, by a factor up to $\sim 3$ during the last $100 \Myr$, which contradicts the assumption of a constant SFR made by \citet{Karl2011} to interpret the observed age distribution of clusters in the Antennae. A longer time-lapse between the pericenter passages would allow the SFR to reach a relatively constant value after the first starburst and would thus be more consistent with the assumption of \citet{Karl2011}. In our simulation, the SFR averaged over the last $100 \Myr$ is higher than the present-day value. Therefore, assuming a constant SFR based on the observed value leads to an underestimate of the dissolution rate of clusters in that age range. This would then support the idea of \citet{Fall2005} and \citet{Karl2011} that the age function of star clusters is mainly due to their dissolution and not their formation rate, but not for the reasons these authors claim. However, the lack of observed older clusters ($\sim 1 \Gyr$) in the Antennae is explained by all models by the low SFR ($\sim 1\Msunyr$) before the first encounter, i.e. by a formation process rather than a disruption one.

\tab{sfr} indicates the stellar mass formed during the three major phases of the merger. Despite their physical differences, the main two starburst episodes produce the same amount of stars, in comparable time-lapses. Furthermore, the phase of separation hosts a non-constant yet continuous formation of stars, producing 1.5 times more stellar mass than each burst, although it does not count any major starburst episode itself. In other words, the interaction produces more stars during the separation than during each individual starburst episode.

\begin{figure}
\includegraphics{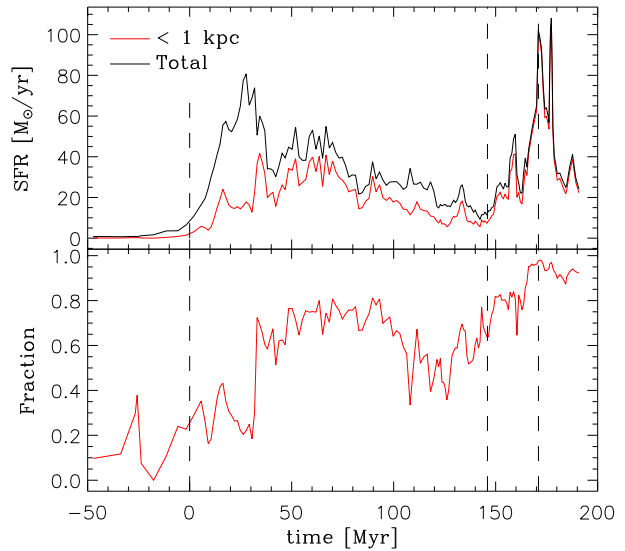}
\caption{Evolution of the SFR during the interaction in the central regions of the galaxies ($< 1 \kpc$), compared to the global SFR. Extended star formation exists throughout the merger, except at the second starburst episode (associated with the third pericenter passage and the final coalescence, $t \gtrsim 170 \Myr$) when it is almost exclusively central, due to gas inflows.}
\label{fig:extended_sf}
\end{figure}

\fig{extended_sf} shows the spatial extent of star formation during the merger. Gravitational torques generate inflow of gas toward the central region throughout the merger. However, as suggested in \citet{Renaud2014a}, extended star formation exists before the pericenter passages themselves ($t = -20 \mh 30 \Myr$ and $t = 100 \mh 150 \Myr$), being triggered by compressive tides and compressive turbulence, in addition to the nuclear inflows. The main starburst occurring at the beginning of final coalescence ($t = 150 \mh 190 \Myr$) is almost exclusively central. Such variations in the spatial distribution of star formation reveal once again the different nature of the two starburst episodes. We explore this point in term of formation of star clusters in \sect{cfr}.

\begin{table}
\caption{Stellar mass formed along the merger}
\label{tab:sfr}
\begin{tabular}{lcc}
\hline
Phase & Duration [Myr] & Mass [$10^9 \Msun$] \\
\hline
First burst (A to B) & 50 & 1.9 \\
Separation (B to C) & 93 & 2.8 \\
Second burst (C to D) & 55 & 1.9 \\
\hline
\end{tabular}
\end{table}

\section{Detection of young stellar objects}
\label{sec:classification}

\begin{figure}
\includegraphics{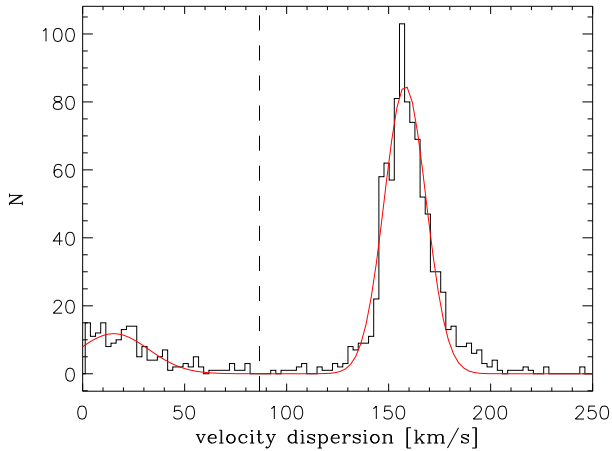}
\caption{Distribution of the velocity dispersion of the stellar objects detected in the simulation. The red line indicates a fit by a sum of two Gaussian functions, with means values of $15.2 \kms$ and $158.1 \kms$. The average of the two means (dashed line, $86.6 \kms$) is the velocity dispersion criterion used to tell appart the two populations.}
\label{fig:veldisp}
\end{figure}

For the analysis presented here, the dense, cluster-like objects formed during the simulation are detected thanks to the friend-of-friend algorithm HOP \citep{Eisenstein1998}, around peaks of local stellar density $> 100 \msun \pc^{-3}$. Sub-clumps are merged if the saddle density between them exceeds $100 \msun \pc^{-3}$ and an outer boundary of $0.1 \msun \pc^{-3}$ is set to tell apart clump members from field stars and tidal debris. For simplicity the stars present in the initial conditions and the gas are not taken into account in the computation of the densities. The output of the detection method provides the membership of all stars to either an overdensity or the field. The properties of the objects, like the mass and the half-mass radius are then derived.

Using a different set of density criteria slightly changes the estimate of the properties of the detected objects, but does not affect our conclusions. Modifying these by an order of magnitude would lead to split some clusters or, on the contrary, to a too severe truncation of viable objects. 

By using a criterion based on the density only, and not involving kinematical information, our method allows us to detect unbound objects. In order to make a difference between genuine star clusters and unbound structures which might be transient over-densities, we adopt additional selection criteria based on the velocity dispersion of the stellar object.

\fig{veldisp} shows the distribution of internal velocity dispersion of all the stellar objects detected by the friend-of-friend algorithm at the final coalescence (epoch D). This velocity dispersion $\sigma$ is the standard deviation of the three-dimensional velocities of the stars contained within the half-mass radius of the object. Two populations appear: the majority (80 per cent) of the objects have a high velocity dispersion, of the order of the galactic-scale motions like the rotation speed of the disc or the relative velocity of the two galaxies, i.e. $> 100 \kms$. The other objects have much lower velocity dispersions. We tell apart the two populations by fitting the global distribution with a sum of two Gaussian functions. The average of the means of the two Gaussians ($\sigma_c = 86.6 \kms$) is then used as a selection criterion.

\begin{figure}
\includegraphics{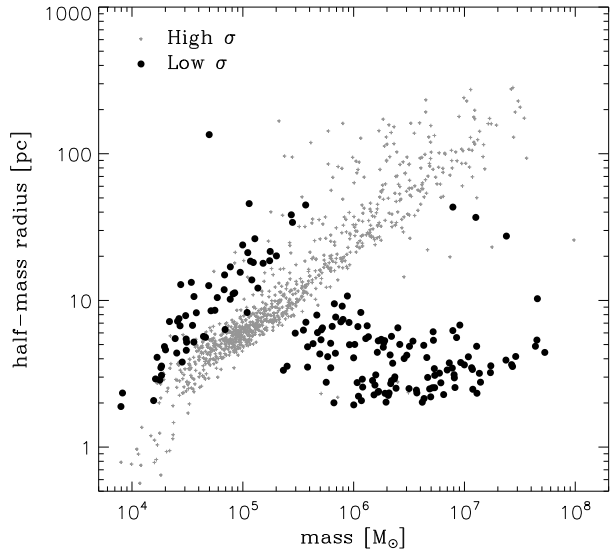}
\caption{Mass-size diagram of the stellar objects detected in the simulation. The small grey symbols correspond to structures with a large velocity dispersion.}
\label{fig:selection}
\end{figure}

In the mass-half mass radius plane ($M - \rh$) shown in \fig{selection}, almost all the objects with a high velocity dispersion ($\sigma > \sigma_c$) lie on a sequence ($\rh \propto M^{0.6}$) running over the full mass range we probe. These objects are transient structures in which the stars enter and leave within a few Myr, in a similar fashion as density waves. Because of the mixing of their content, these structures count stars with a wide diversity of ages, that have formed under different conditions and environments, populate the field, and eventually are temporarily gathered in such over-densities.

\begin{figure}
\includegraphics{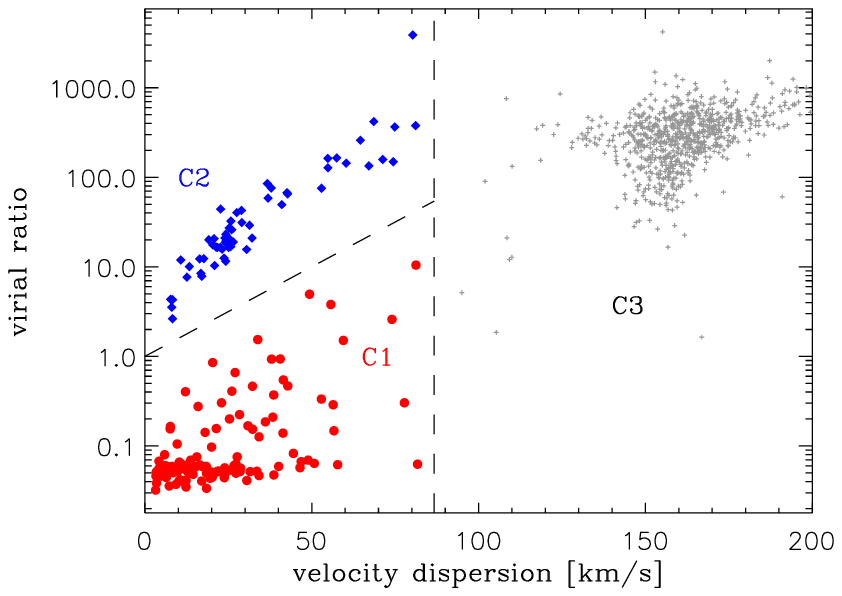}
\includegraphics{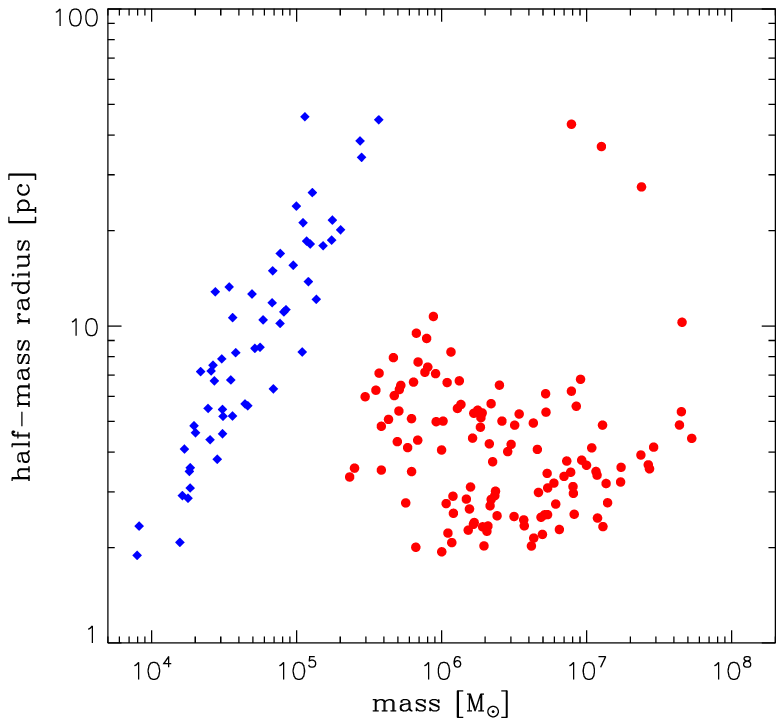}
\caption{Top: virial ratio ($\sigma^2 \rh / GM$) plotted against velocity dispersion ($\sigma$) for all the stellar objects detected in the simulation. After removing the objects with high velocity dispersions, we split the remaining into two classes and use this identification in the mass-size diagram of the bottom panel where they are still well-separated.}
\label{fig:virial}
\end{figure}

By estimating the dimensionless virial ratio ($\sigma^2 \rh / GM$), we can split the objects with low velocity dispersion into two families (\fig{virial}). This also corresponds to the separation visible in the mass-size diagram, on both sides of the sequence, as noted above. Note that the softening of the gravitational potential at small scale (on the resolution of the grid, i.e. $1.5 \pc$) introduces a bias tending to stabilize small stellar structures against collapse. For these reasons, we over-estimate the potential energy and thus, underestimate the virial ratio of the small objects. Nevertheless, the separation made in \fig{virial} tells apart bound objects (class C1, low virial ratio) from unbound ones (class C2), even at small radii, on top of the separation from structures with a high velocity dispersion (class C3) already made.

\begin{figure}
\begin{center}
\includegraphics{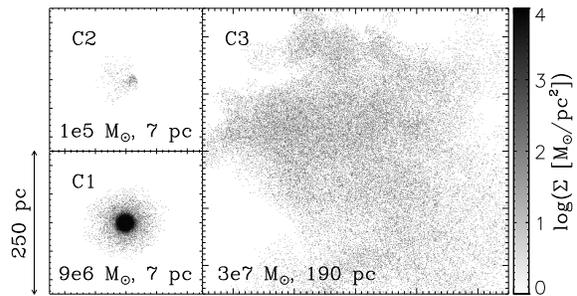}
\end{center}
\caption{Surface density of stellar objects illustrative of the three classes defined in \fig{virial}: low velocity dispersion and low virial ratio (bottom left, class C1), low velocity dispersion and high virial ratio (top left, C2) and high velocity dispersion (right, C3). The mass and half-mass radius of the object are indicated in each panel. The scalings (size, bin size and color) are the same for all three panels.}
\label{fig:clus_morpho}
\end{figure}

\begin{figure}
\begin{center}
\includegraphics{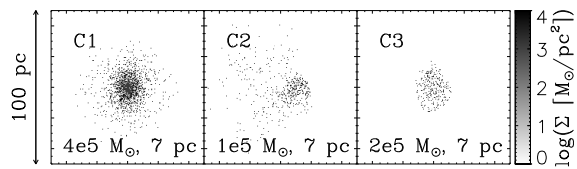}
\end{center}
\caption{Same as \fig{clus_morpho} but for three objects with the same half-mass radius, and comparable masses. The object from the C2 class is the same as the one shown in \fig{clus_morpho}.}
\label{fig:clus_morpho2}
\end{figure}

\fig{clus_morpho} and \fig{clus_morpho2} illustrate the differences in morphology between these three classes of objects. Even if all are over-densities of young stars, only the C1 objects ressemble star clusters, with a clear central peak density and a spherical shape. The C2 class seems equivalent to (massive) associations, while the C3 structures are transient over-densities of field stars, with no particular features or sub-structures\footnote{The typical surface density and velocity dispersion of C3 objects are of the same order of magnitude than that of galactic structures.}. The classes C1 and C2 represent well the diversity of binding of stellar objects observed by e.g. \citet{Maiz2001}. A more detailed comparison to observed objects is made in \sect{obs}. In the following, we will refer to C1 objects as star clusters and to C2 objects as associations.

The bi-modality in velocity dispersion does not exist throughout the interaction: during the earliest phases (before and during the first encounter), almost all the objects detected have a velocity dispersion smaller than $20 \kms$. The class C3 does not exist at these epochs. After the first starburst, the bi-modality appears and the high velocity dispersion mode slowly shifts towards higher and higher values. Therefore, we adjust the value of our criterion $\sigma_c$ as a function of time to separate the C3 structures form the C1 and C2 objects.

\begin{figure}
\includegraphics{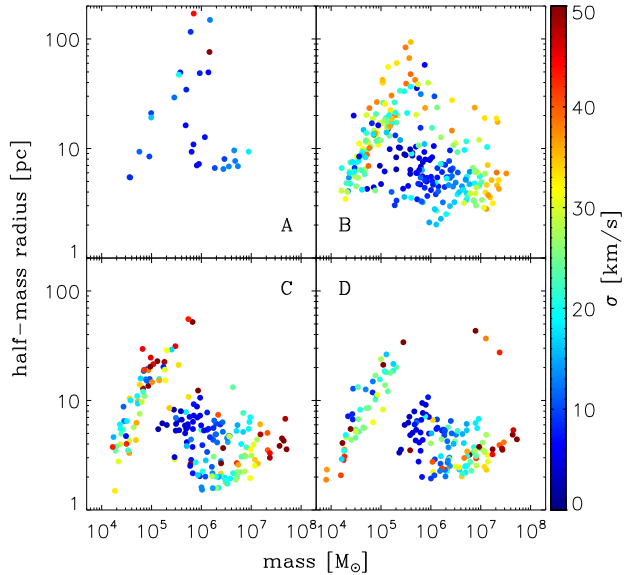}
\caption{Mass-size diagram of the star clusters and associations at the epochs A, B, C and D marked in \fig{sfr}. The colour represents the velocity dispersion.}
\label{fig:mr_evol}
\end{figure}

\fig{mr_evol} shows the distribution of star clusters and associations in the mass-size plane at the four epochs highlighted in \fig{morphology} and \ref{fig:sfr}, with colour indicating the velocity dispersion. At all epochs, we detect sub- and super-virial objects (classes C1 and C2). These two classes are well separated most of the time (using the same criterion than in \fig{virial}, at the epoch D), but the distinction is less clear during the first starburst (pictured here at epoch B). At this epoch, the extreme conditions of the galactic collision in term of SFR and tidal effects probably create objects with intermediate virial ratios and/or shift existing objects to this regime. Evolution processes make them collapse, expand or even dissolve and thus re-creating the clear cut between the C1 and C2 classes. Since the criterion on the virial ratio adopted at final coalescence applies also during earlier stages, we choose to use it for all snapshots.

In the rest of the paper, we focus the analysis and discussion on star clusters (C1).

\section{Formation and early evolution of star clusters}
\label{sec:formation}

\subsection{Cluster formation rate}
\label{sec:cfr}

\begin{figure}
\includegraphics{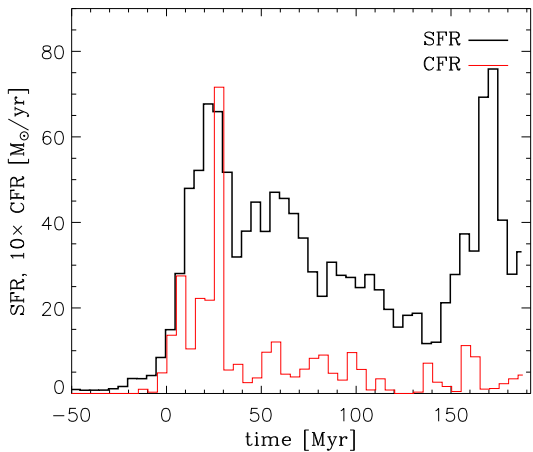}
\caption{Cluster formation rate (multiplied by 10, for the sake of clarity), compared to the star formation rate, as functions of time. Both quantities are computed in bins of $5 \Myr$, which explains the differences in absolute values with that of the \fig{sfr}. Only C1 objects are considered.}
\label{fig:cfr}
\end{figure}

Using the detection and diagnostics presented in \sect{classification}, we identify the star clusters (C1 only) on snapshots every $\sim 5 \Myr$. The time distribution of the first detections of clusters then gives us the cluster formation rate (CFR), shown in \fig{cfr}, and compared to the SFR recomputed with the same time binning (i.e. $5 \Myr$). Obviously the CFR only traces the mass of clusters at the epoch of their formation, ignoring their future growth or evaporation. As discussed in \sect{sfr}, after $t \approx 45 \Myr$, star formation is not active anymore in the tidal debris, but limited to the central parts of the discs. The CFR suddenly drops by a factor $\sim 15$ showing indeed that star formation only proceeds in already formed clusters, and in the galactic nuclei. This remains valid during the separation of the galaxies, and even for the second major starburst episode ($t \approx 150 \Myr$) meaning that the vast majority of clusters are formed by the first galactic encounter.

\begin{figure}
\includegraphics{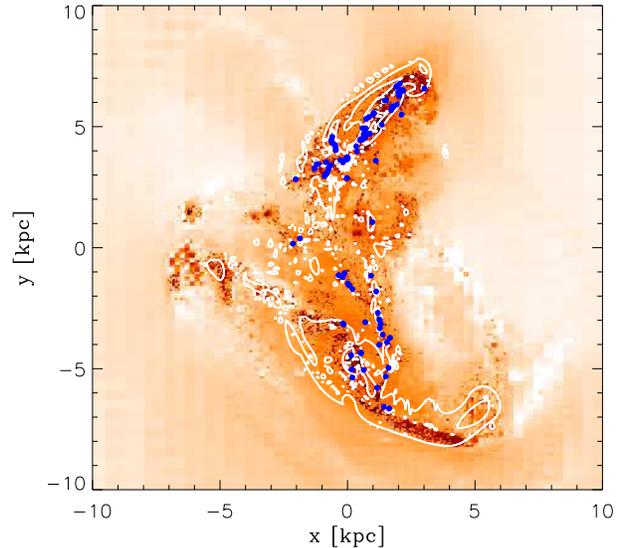}
\caption{Correspondance between compressive tides, compressive turbulence and cluster formation during the first starburst episode ($t = 20 \Myr$). The colour map represents the strength of the tidal field (computed as the density weighted average of the largest eigenvalue of the tidal tensor at the scale of $40 \pc$, see \citealt{Renaud2014a}), running from no tides (white) to strongly compressive tides (red). Contours mark the regions experiencing converging compressive turbulent flows, and the blue dots indicate the position of the young star clusters ($< 10 \Myr$).}
\label{fig:compressive}
\end{figure}

\citet{Renaud2014a} established a relation between compressive tides the importance of which significantly rises in mergers, compressive turbulence and star formation. \fig{compressive} illustrates the spatial correspondance between the three aspects, short after the first pericenter passage.\footnote{A detailed study of the evolution of this correspondance showing the propagation of compressive tidal waves will be presented in \citealtip{\Bournaud}} Although a delay of $\sim 20 \Myr$ exists between the initial tidal trigger and its effect on the star formation rate, we confirm that almost all young star clusters are found inside tidally and turbulently compressive regions. Such large-scale effects trigger star (cluster) formation over spatially extended volumes, and not only in the galactic nuclei. However, the shorter impact parameters of the second and third encounters imply a more important role of gas inflow onto the nuclei. Therefore, star formation occurs mostly in the nuclei and not in a number of star clusters at various distances from the centre as it does during the first pericenter passages. At $t = 170 \Myr$, 82 per cent of the young stars ($< 2 \Myr$) form in the central $500 \pc$ of the merger, in the galactic nuclei, and thus do not form new clusters: the CFR remains low despite a burst of the SFR.

Note that the non-equality between the CFR and SFR does not invalidate the fact that all stars form in clusters: in our simulation, stars can form in pre-existing clusters (or in the galactic nuclei), such that the SFR remains high despite a zero or small CFR.

\begin{figure*}
\includegraphics{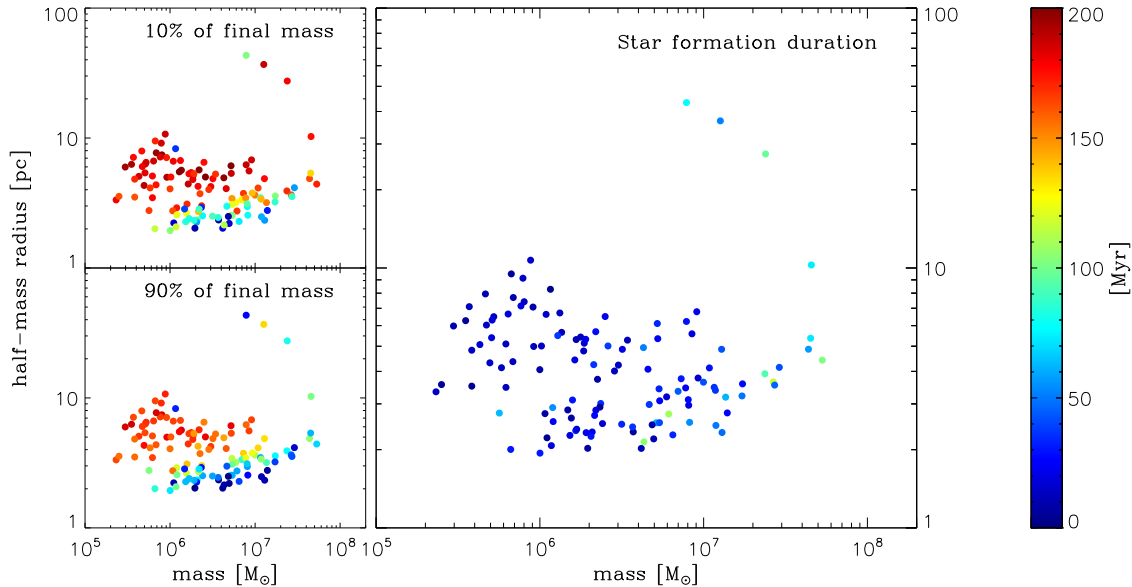}
\caption{Mass-size diagram of the star clusters (C1 only) formed during the simulation, and detected at the final coalescence (epoch D, $t= 190 \Myr$). The colour indicates the time difference between this instant and the moment when the cluster had formed 10 per cent of its final stellar mass, i.e. about the age of the oldest stars (top-left), and 90 per cent of the final mass, i.e. about the age of the youngest stars (bottom-left, see text). The right panel shows the difference between the two, i.e. an indicator of the duration of the star forming episode.}
\label{fig:mr_duration}
\end{figure*}

To further investigate the formation of clusters, we define three quantities: (1) the epoch when a cluster has formed 10 per cent of its final mass, i.e. a proxy for the beginning of its star forming episode, (2) the epoch when it has formed 90 per cent of its mass, i.e. a proxy for the end of its star forming episode, and (3) the difference of the two, i.e. the duration of the star forming episode. Setting our thresholds to 10 and 90 per cent instead of 0 and 100 avoids the contamination by isolated star forming episode in the lowest mass clusters, where the star formation history is sensible to Poisson's noise.

\fig{mr_duration} displays these three quantities for the clusters detected at the epoch D. The absence of relation between the mass and the epoch of formation reveals that massive objects are formed throughout the interaction process. It however exists an age-half mass radius relation: the most extended object formed earlier. This could indicate either an evolution pattern, namely the expansion of the clusters, or general properties of the cluster they maintain since their birth. \fig{mr_evol} reveal that in the early stages of the merger, young clusters can be detected with a relatively large radius ($\sim 10 \pc$) before any internally-driven evolution takes place\footnote{The Gyr-long evolution of such clusters is expected to drive them through an expansion phase \citep{Gieles2011b}, even taking the time-dependent tidal field they experience into account \citep{Renaud2013a}. This effect is negligible over the timescales considered here.}. Therefore, the extended nature of the clusters formed early is a signature of the conditions of their formation. The interaction and the associated compression by tides and turbulence \citep{Renaud2014a} tend to form dense but small gas clumps. The star clusters formed in these clumps inherit their physical properties. Clusters formed before a pericenter passage are born in a less fragmented ISM than those born after: the fraction of dense gas ($\gtrsim 10^4 \cc$) is multiplied by a factor 5 to 10 by the interaction, as visible in the density probability distribution functions of \citet[their Fig. 2]{Renaud2014a}.

The mean duration of the star forming episode is $40 \Myr$, but the distribution peaks at $10 \Myr$ indicating that most stellar objects have formed during a single, short episode and are thus coeval. A few clusters however yields longer star formation durations, up to $100 \Myr$. Such distribution of star formation histories is compatible with the estimates of \citet{Peterson2009} who observed the central regions and the tidal features of the interacting galactic pair Arp 284 and found a majority of short formation episodes and only a few $\sim 100 \Myr$-long cases. Our findings however contradicts the observations of YMCs in the merger remnant NGC~343 by \citet[see also \citealt{Bastian2013b}]{Cabrera2014} who advocate a single, short ($\lesssim 2 \Myr$) formation event. As proposed by \citet{Peterson2009}, the difference of galactic environment, which in these cases is mostly due to different merger stages\footnote{Arp 284 is likely to be observed $\sim 100 \Myr$ after the first pericenter passage and before the second one \citep{Struck2003}, while NGC~34 is in the coalescence phase.} could play a role on the star formation histories of clusters.

In our simulation, long, extended star formation histories are split into several star formation episodes separated by $\sim 50\mh 80 \Myr$. The episodes producing most of the stars in a given cluster are associated with the first and second galactic pericenter passages. (The details of this timing vary depending on the exact position of the clusters in the galaxies when they collide: clusters in the outermost regions of the galaxies form their second stellar population before the galactic pericenter passage itself.) Therefore, the interaction acts as a global trigger to form star clusters, but also their secondary stellar populations.

We also note that the most massive clusters gain a fraction of their mass by accreting smaller objects. Since these smaller objects also preferentially formed during the global starburst, the resulting stellar population in the final cluster yields a small spread in age, shorter than the duration of the starburst, but not two distinct populations. An analysis of the chemical abundances of such objects (impossible in our simulation) may unveil the complexity of their star formation histories. 

As mentioned before, computing the gravitational potential on a grid (particle-mesh method) and softening the gravitational potential implies that the collisional aspect of the star clusters and thus, the full dynamical evolution cannot be probed accurately. Simulations accounting for this do exist but the price for accuracy is a strong limitation in the number of bodies (i.e. stars) included. Simulating the $N$-body driven evolution of several $10^5 \Msun$ collisional objects remains therefore out of reach. One can argue that such evolution for clusters in the mass range we consider is very slow compared to the duration of the merger, and thus can be neglected in a first approach. (Core-collapse for such clusters would take $\gtrsim 1 \Gyr$, see \citealt{Alexander2012}.) Other drivers, related to $\lesssim 100 \Myr$ long evolution, are accounted for in our simulation, at least to some extent (gas accretion, tides, relaxation).

\subsection{Comparison with the Milky Way simulation}

\begin{figure}
\includegraphics{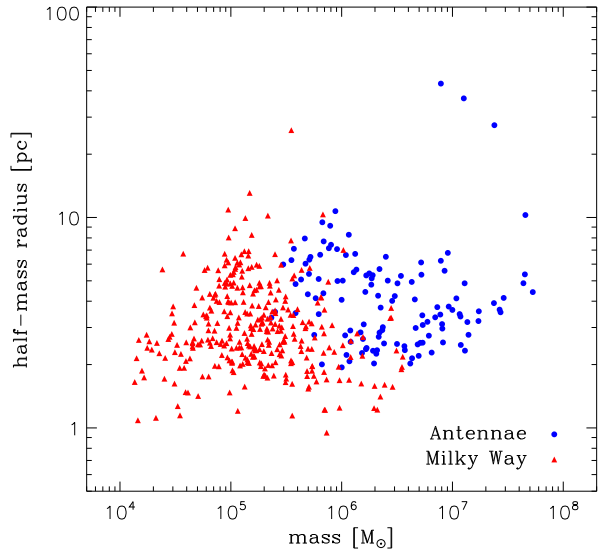}
\caption{Mass-size diagram of the star clusters detected in the simulations of the Antennae and of the Milky Way.}
\label{fig:mr_mw}
\end{figure}

The same detection method and analysis have been applied to the MW simulation \citep{Renaud2013b}. At the instant chosen, the MW counts 341 star clusters (C1), and 21 associations (C2). All objects yield velocity dispersion below $30 \kms$ (the mean value being $9 \kms$).

\fig{mr_mw} compares the distributions of the star clusters in the simulations of the MW and the Antennae in the mass-size diagram. We found clusters spanning several decades in mass and $\sim 1.5$ dex in half-mass radius, with no correlation between the two quantities, and mostly overlapping with the loci of the observed Milky Way clusters ($M \sim 2\times 10^5 \Msun$, $\rh \sim 3 \pc$, see \sect{obs}). Furthermore, no global correlation is found between the mass (or size) of the clusters and their position within the simulated disc. However, the objects with the highest velocity dispersion are found close to the tip of the bar, and in the portion of spiral arm undergoing fragmentation.

Overall, although the cluster populations of the two galaxy models are comparable in many aspects, we note several divergences. First, contrary to the Antennae, the MW simulation does not contain very massive clusters ($\gtrsim 4\times 10^6 \Msun$). Furthermore, the average mass of clusters in the Antennae is 15 times that in the MW. This point is further discussed in \sect{cmf}. Second, the Antennae clusters are generally more extended than those of the MW (by a factor 2, in average). In the end, the cluster in the merger are $\sim 2.6$ times denser than in the MW. Third, none of the MW clusters is found in massive and extended regime ($\sim 10^7 \Msun$, $\sim 30 \pc$) whereas an handful of cases exists in the Antennae, clearly separated from the bulk of the distributions. This reflects the effects of the rapidly varying environment of these clusters (see \sect{w3}).


\section{Cluster mass function}
\label{sec:cmf}

\subsection{Evolution of the CMF}

\begin{figure}
\includegraphics{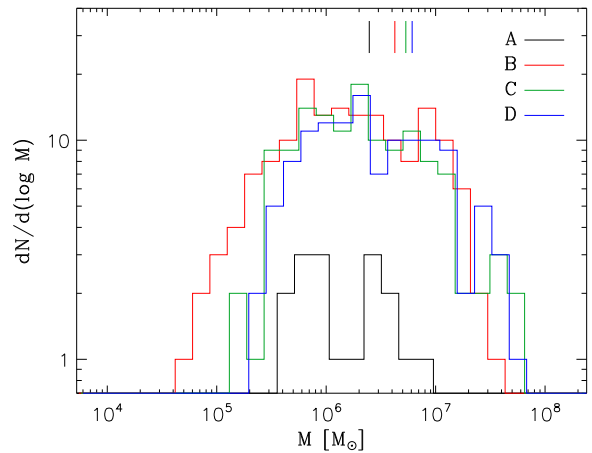}
\caption{Cluster mass function in logarithmic bins of mass, at the four epochs A, B, C and D marked in \fig{sfr}. Vertical markers indicate the average mass of the clusters for each distribution. (The histograms are slightly shifted horizontally with respect to each other, for the sake of clarity.)}
\label{fig:cmf}
\end{figure} 

\fig{cmf} shows the cluster mass function (CMF) at the four stages marked in \fig{sfr}. In the Antennae, before the first SFR peak (epoch A), the long-range interaction has already triggered the formation of a handful of clusters.

At the first post-starburst epoch (B), following the peak of CFR (recall \fig{cfr}), the number of clusters has been multiplied by $\sim 15$. The tidally triggered fragmentation of the ISM leads to the formation of many  $\lesssim 4\times 10^5 \Msun$ clusters and also very massive cases, up to $\sim 5 \times 10^7 \Msun$. The average mass of clusters doubles with respect to the isolation stage, as noted observationally by \citet{Larsen2009}. At the end of the separation of the galaxies (epoch C), the CMF is comparable to that of the previous stage for $M \gtrsim 10^5 \Msun$ but the number of low-mass objects decreases. A deficit of intermediate mass clusters ($\sim 10^5 \Msun$) appears. Finally, during the second starburst and at the final coalescence (epoch D), a few very massive (up to $\sim 10^8 \Msun$) objects are detected in the central region of the merger, increasing the average cluster mass by a factor 1.8. The deficit of intermediate-mass clusters is conserved. Such deficit leads to a major difference between bell-shaped simulated CMF and the observed one \citep[see e.g.][who detected a power-law shape down to $\sim 10^4 \Msun$]{Whitmore2010}. Since our numerical method allows for the formation of intermediate mass clusters (see the next Section), we do not think that the difference principally arises from a numerical artefact, but rather from physical conditions. It is however possible that interaction with dark matter particles could destroy small clusters. The mass ratio between dark matter and stellar particles is $\sim 2400$ in our simulation. Such a high ratio could induce strong effects due to two-body interaction \citep{Steinmetz1997}, in a comparable fashion as the destruction of low-mass clusters by collisions with giant molecular clouds \citep{Gieles2006c}. Unfortunately, decreasing this mass ratio would require a very large number of dark matter particles ($\sim 2\times 10^9$) and thus a much higher numerical cost, which prevents us to quantity this effect.

\subsection{Comparison with the Milky Way simulation}

\begin{figure}
\includegraphics{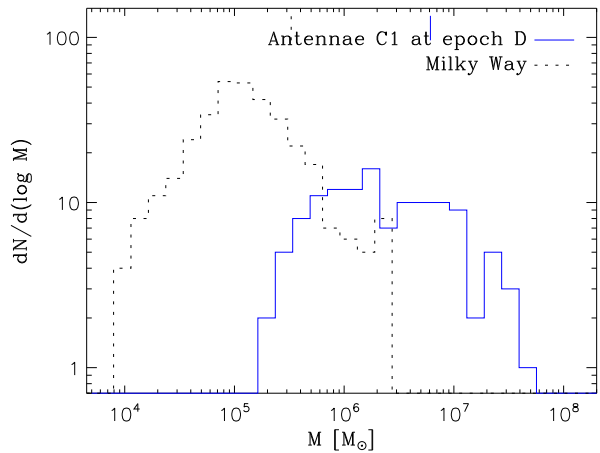}
\caption{Cluster mass function in the Antennae and the Milky Way simulations.}
\label{fig:cmf_mw}
\end{figure} 

The CMF in the Milky Way model yields a log-normal shape (see \fig{cmf_mw}). The most massive cluster in the MW simulation has a mass of $3\times 10^6 \Msun$ and is found at the tip of the galactic bar, where gas clouds in the bar interact with those from the spirals arms to form giant molecular associations and massive young clusters. Cloud-cloud collisions are known to trigger intense, localized episodes of star formation in such environment (\citealt{Motte2014}, see also \citealt{Furukawa2009, Fukui2014} about collisions in other environments). 

Cloud-cloud collisions are also expected in galaxy mergers, in particular at the coalescence phase. Unfortunately, the Eulerian nature of our simulation makes it difficult to follow the trajectories of gas clouds and estimate their collision rate, in particular in a fast-evolving systems like the Antennae. Without being able to quantify precisely their contribution to the formation of clusters, we note that the cloud-cloud collisions provide sufficient pressure to trigger the formation of massive clusters (\citealt{Furukawa2009}, \citealt{Elmegreen2014}, see also \citealt{Elmegreen2001c}), and such conditions are found in some specific areas like the overlap region of the merger. Other factors like clouds being shocked by a density wave created by the galactic interaction and the compressive trigger introduced in \citet{Renaud2014a} also favor the formation of massive clouds and subsequently massive clusters. All together, these processes explain the difference in the maximum mass of star clusters between an isolated galaxy and a merger.


\section{Variety of dense stellar systems}
\label{sec:variety}

\subsection{Simulated and observed structural properties}
\label{sec:obs}

\begin{figure*}
\includegraphics{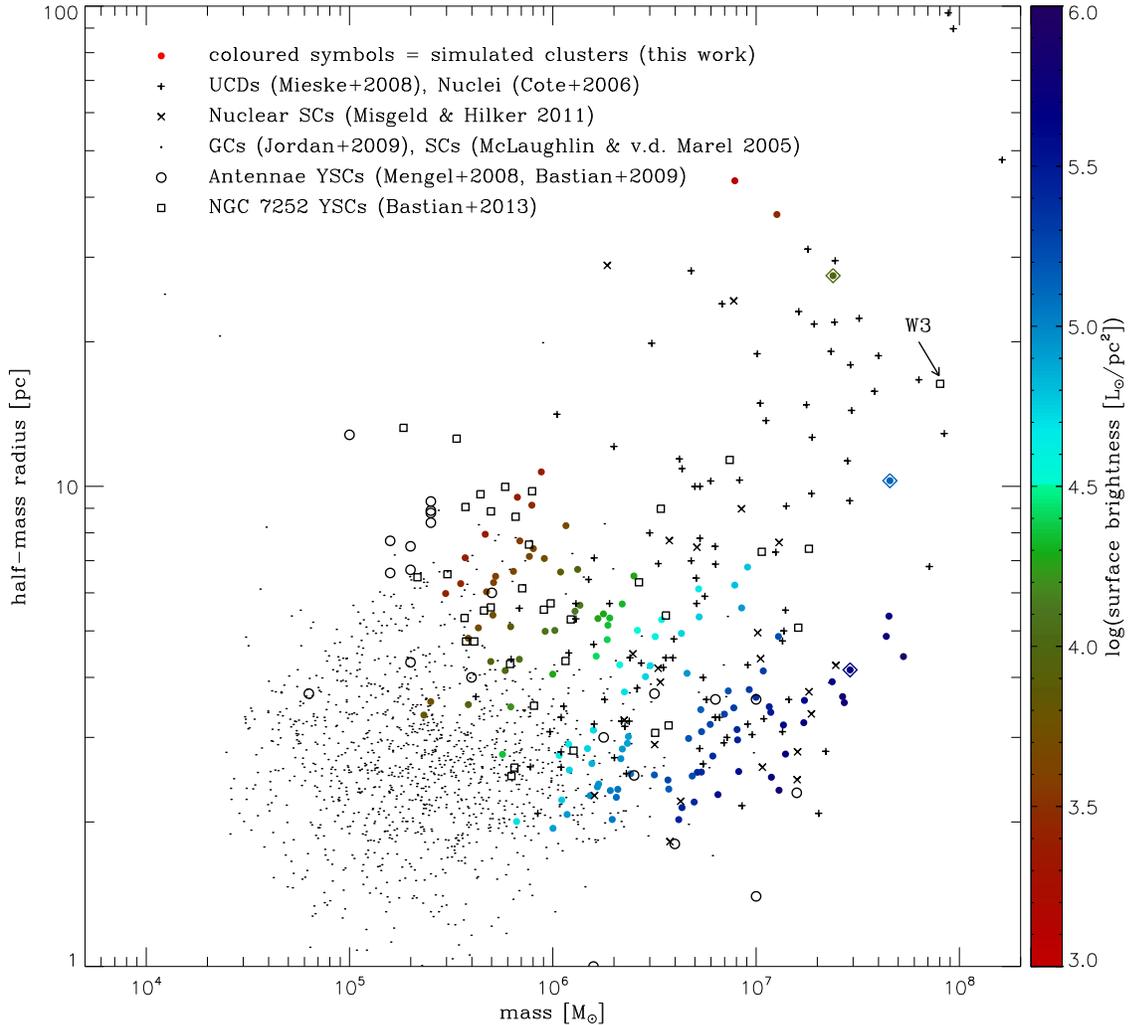}
\caption{Mass-size diagram of the stellar objects formed during the simulation and detected at the final coalescence (epoch D), shown with coloured dots. The colour represents the surface brightness of the cluster, computed from \eqn{s}. Clusters in diamonds are discussed in Sections~\ref{sec:w3} and \ref{sec:tails}. Black symbols refer to observations of dwarf galaxies and star clusters in the local Universe.}
\label{fig:mr_obs}
\end{figure*}
\nocite{Misgeld2011, Mieske2008, Cote2006, Jordan2009, McLaughlin2005, Bastian2009, Mengel2008, Bastian2013}

A full comparison between the simulation and the real Antennae galaxies is presented in \citetip{\Duc}. In this Section, we use the simulation as a prototype of Antennae-like mergers, without seeking a one-to-one correspondance with the Antennae themselves. The analysis is made using the data at the end of the simulation, at coalescence (Epoch D), i.e in the post-merger phase when it is most probable to observe interacting galaxies (owing the short duration of the collision phases).

During the close encounters of the galaxies, where the observational classification of systems as mergers is the most robust, star forming regions are generally strongly obscured by the overlap of the ISM of two galaxies. Therefore, only dense, young and bright clusters are observationally detected. Keeping this issue in mind, we analyse the resemblances and differences between our simulated objects and the compilation of observations from \citet{Misgeld2011}, to which we add published data specific to the Antennae \citep{Mengel2008, Bastian2009} and to NGC~7252 \citep[a merger remnant,][]{Bastian2013}. \fig{mr_obs} shows the positions of the observed and simulated stellar objects in the mass-size plane, assuming that the observed effective radius corresponds to the simulated half-mass radius.

Assuming that each stellar particle of the simulation fully samples the IMF and represents, by definition, a coeval population, we evaluate its luminosity in the V band as a function of its mass $m$ and age $a$:
\begin{equation}
L_{V\star} = L_{V\odot} \frac{m}{1 \Msun} 10^{-0.4 (a / 1 \Myr)^{-0.7}},
\label{eqn:lum}
\end{equation}
following \citet{Boutloukos2003}. We compute the intrinsic surface brightness of a cluster $S_{V,c}$ by summing the luminosities of its members and using the surface given by the half-mass radius:
\begin{equation}
S_{V,c} = \frac{\sum_{i} L_{V\star,i}}{\pi \rh^2}.
\label{eqn:s}
\end{equation}

The simulated star clusters lie in the same loci in the mass-size diagram than the young massive clusters observed in the Antennae (\fig{mr_obs}). The most massive cases are comparable to clusters detected in the overlap region ($\sim 10^7 \Msun$, $\sim 3 \pc$, \citealt{Mengel2008}). \fig{mr_evol} also tells us these clusters have been the last ones to form, in agreement with observational data. The least massive end of the distribution of simulated clusters is comparable to the clusters detected in more external regions of the galaxies ($\sim 2\times 10^5 \Msun$, $\sim 8 \pc$, \citealt{Bastian2009}), compatible with an earlier formation. The distribution is also compatible with that of the clusters found in another merger, NGC~7252, by \citet{Bastian2013}.

Our clusters are offset from the bulk of the Milky Way and extragalactic globular clusters ($\sim 2\times 10^5 \Msun$, $3 \pc$), in both the observations and the simulations. The simulation reproduces the global effect of a merger in shifting the distribution of star clusters toward more massive and slightly larger objects for the reasons discussed in \sect{cmf}. Furthermore, the different physical conditions between the two starbursts (tidal compression for the first one, and gas inflow for the second, see \sect{cfr} and \citealt{Renaud2014a}) explain the spread of the distributions, from intermediate-mass, extended clusters formed early, and massive compact object formed late.

The most extended objects in the high-mass half of our distribution are in the regime of ultra-compact dwarf galaxies (UCDs) and nuclear star clusters. The formation scenario and the properties of two of them are discussed in details in \sect{w3}.

We note that the maximum mass of objects found in the simulation ($\sim 10^8 \Msun$) is higher than that detected in the real Antennae galaxies, but is in good agreement with objects observed in a wider sample of environments \citep{Norris2014}. The difference between our model and the Antennae is therefore a lack of realism either in the physics implemented, and/or in the realism of the model itself (initial conditions, orbit, etc.). We expect the sub-grid physics and the neglected physical phenomena like magnetic field and two-body relaxation to affect mostly the objects made of a few stellar particles but to have a milder impact on the global properties (mass, size) of the largest cases over the timescales considered here. Interestingly enough, the physical setup leading an Antennae-like system allows to produce most of the $\lesssim 100 \pc$ stellar objects observed, including an handful of the most massive ones (see next Section). This suggests that real systems showing a small variation in their parameters with respect to the Antennae galaxies could host the formation of massive objects, comparable to those seen in our simulation.

\subsection{Formation of ultra compact dwarf galaxies}
\label{sec:w3}

Among the dense stellar objects, ultra-compact dwarf galaxies (UCDs) are at the frontier between star clusters and galaxies, such that their formation scenario is still debated. It has been proposed that UCDs could be the remnant of a tidally-stripped dwarf galaxy \citep[see e.g.][]{Zinnecker1988, Norris2014}. However, \citet{Mieske2012} argued that these objects lie in the most luminous and massive ends of the distributions of globular clusters which indicate the two families of objects share the same formation mechanism (see also \citealt{Caso2014}, Zhang et al. in prep and Liu et al. in prep. which propose opposite conclusions).

The W3 stellar object, observed in the galaxy merger NCG~7252 and marked in \fig{mr_obs} is found between the regimes of the massive star clusters and that of the ultra-compact dwarf (UCD) galaxies ($M \approx 8 \times 10^7 \Msun$, $\rh \approx 17 \pc$, \citealt{Maraston2004, Fellhauer2005}). Using numerical simulations, \citet{Fellhauer2005} proposed that W3 could form through merging of star clusters formed during a starburst episode.

 These authors successfully found a set of parameters (structural and orbital) for a star cluster complex made of sub-clusters to reproduce the observed characteristics of W3 (with however a discrepancy between the modelled luminosity profile and the observations, see \citealt{Bastian2013}), arguing that such configurations are commonly found in interacting galaxies. Our galaxy simulation can be used to complement their star cluster simulation and discuss the likelihood of their scenario. 

At the end of our simulation, several objects share the properties of observed UCDs, and two in particular are comparable to W3. The first has a similar mass but is less extended ($M \approx 5 \times 10^7 \Msun$, $\rh \approx 10 \pc$). The gas cloud leading to this cluster forms in the outer region of NGC~4039 facing NGC~4038, along with many clumps of comparable mass distributed along a $\sim 2 \kpc$ long gaseous arm which fragments under the effects of the tidal interaction. Because of the rotation of the galactic disc, this region rapidly moves to the other side of the system (within $10 \Myr$, at $\sim 500 \kms$ in the global reference frame), close to but not within the tidal tail. The gas clumps remain relatively close to the galaxy center ($\lesssim 4 \kpc$). During this process, the gas clouds gather in a small volume, in a comparable fashion as the piling-up scenario proposed in \citet{Duc2004}, and rapidly merge into a handful of clumps. Our cloud of interest (that will form the W3 analogue) accretes gas from its host arm and form stars at $\sim 1.5 \Msunyr$ during $\approx 30 \Myr$. Star formation stops when the surrounding interstellar medium becomes too diffuse to continue to fuel gas accretion onto the cloud. The young cluster then keeps a constant mass, and its orbit keeps it close to the galactic center (within $2 \pm 1 \kpc$) during the entire simulation.

\begin{figure}
\includegraphics{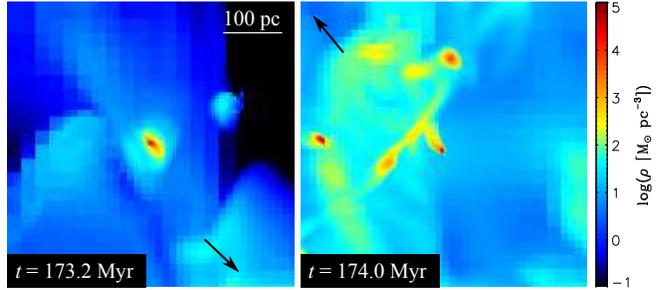}
\caption{Density maps ($500 \pc \times 500 \pc $) of the gas cloud associated with our second W3 analogue, a few $10^5 \yr$ before (left) and after (right) it flies through the galaxy merger. Arrows indicate the direction of the galactic centre toward which (left) and from which (right) the cloud is moving. Within the $0.8 \Myr$ between these snapshots, the outer envelope of the cloud has been stripped by ram pressure and tides. The inner region has not been strongly affected, but the stellar counterpart starts to re-adjust to the new potential well with an increase of its half-mass radius by a factor 7 within the next $30 \Myr$.}
\label{fig:w3_c5}
\end{figure}

Our second analogue is slightly less massive but more extended ($M \approx  2 \times 10^7 \Msun$, $\rh \approx 27 \pc$). It forms in a comparable environment than the first candidate, i.e. in between the galaxies, but in the other progenitor. It also moves to the other side and remains fairly close to the galactic centre ($4 \kpc$). Its gaseous nursery travels alternatively through dense and diffuse medium and experiences sequentially ram pressure and accretion of gas. Through this sequence of events, the stellar object builds up during $90 \Myr$ (with a SFR peaking twice at $\sim 0.7 \Msunyr$ during the first global starburst). Then, its mass remains approximatively constant until the end of the simulation. This object is compact during most of its life ($\rh \approx 4 \pc$). During the third pericenter passage however, it travels through the central kpc of the merger. The outer layers of the associated gas cloud (i.e. between $5$ and $20 \pc$) are significantly altered: ram pressure and (at a smaller extend) tidal effects strip half of the gas mass within $10 \Myr$ (see \fig{w3_c5}). However, the inner part (within the half-mass mass radius of $\sim 5 \pc$) is dense enough ($1.5 \times 10^6 \cc$), not dynamically affected and conserves its mass and density. After the crossing of the galaxy, the gravitational potential of the cloud is twice as shallow than before, and since the stellar object keeps approximately the same velocity dispersion, it becomes super-virialized (see \citealt{Boily2003a, Boily2003b} for comparable effects due to gas expulsion by stellar feedback). Thus, the cluster grows shortly after the crossing of the merger and its half-mass radius jumps from $4$ to $27 \pc$ within a half-mass relaxation time ($\sim 30 \Myr$). Despite this rapid change, the cluster remains bound and survives.

Both our W3 analogues are free of dark matter and form in the outskirts to the discs, almost in the tidal bridges, during the first starburst, in a somewhat similar fashion than the hinge clumps detected by \citet{Smith2014}. With the analysis of the evolution of the CMF presented in \sect{cmf}, we conclude that the conditions to form such massive and compact objects are only fulfilled during the starburst phases of a galaxy merger. The two formation histories we have highlighted above ($30 \Myr$-long accretion of star forming clouds and violent relaxation of a dense object flying through the center of a galaxy) favors the idea of W3-like objects in particular and UCDs in general being the massive end of the distribution of star clusters and not tidally-stripped dwarf. Stripping is however advocated in the second scenario, but on the gaseous left-overs of a young massive cluster and not on a dwarf galaxy.

The early merger of clouds is comparable to the scenario proposed by \citet{Fellhauer2005}, but we note that the clouds are not part of a single structure or cloud-complex. The hierarchical flavor of their scenario (following \citealt{Bonnell2003}) is not found in our simulation. We note that the numerical dissipation of turbulence in our simulation at the parsec-scale could modify the way the ISM fragments at these scales and bias our findings. However, the hierarchical formation of a massive cluster has been detected elsewhere in the system (see \sect{tails}), indicating that the simulation allows such process to happen.

In conclusion, Antennae-like mergers provide the physical condition for the formation of young stellar objects as massive as W3 and the most compact cases of UCDs. Their formation channel is comparable to that of massive clusters. The age distribution of their stars is extended ($30$ or $90 \Myr$) but continuous. Yet, an additional process is needed to eject the UDCs in intergalactic regions to account for their observed spatial distribution. Future galactic collisions, which are likely to happen in galaxy cluster environment but much less common in galactic pairs, could play this role.

\subsection{A globular cluster in the tidal tails}
\label{sec:tails}

In our simulation, the Southern tidal tail does not contain any clustering of newly formed stars, but an elongated, rather diffuse distribution of a few field stars along the tail, over $35 \kpc$. All these stars formed several $10^7 \yr$ before the first collision, i.e. before the tidal triggering of star formation and the formation of the tail itself. The cluster in which they have been formed has been stripped and a fraction of its stars has been tidally ejected into the tail. The gas content of the tail is too diffuse to be converted into stars and thus, it does not host \emph{in situ} star formation.

Oppositely, the Northern tail does not show spare young stars but a single concentration into a dense stellar object, situated $32 \kpc$ away from the merger remnant (at epoch D). This massive object ($3\times 10^7 \Msun$) is much more compact ($\rh = 4 \pc$) than classical tidal dwarf galaxies (TDGs, \citealt{Duc2014}), although one would expect that the weak tidal field in this region would allow for a higher radial expansion before reaching tidal truncation. However, the duration of the simulation is too short compared to the expansion timescale of such a stellar system, and the tidal field is not strong enough, or has not acted long enough to have significantly heated the stars of this object and impacted its structure \citep[see][]{Renaud2013a}. The stellar population of this object is $\approx 160 \Myr$ ($\pm 20 \Myr$) younger than the tidal tail itself, indicating an \emph{in situ} formation. Furthermore, the spin velocity ($\sim 100 \kms$) of this object is about five times higher than its velocity dispersion, indicating a net rotation, while other objects in the simulation show much less clear signatures of rotation, if any. The presence of such object is compatible with the observations of the Antennae, showing that, although a tidal dwarf galaxy could not form, parsec-size stellar object formed \emph{in situ} exists in the tidal tails \citep{Bournaud2004}.

\begin{figure}
\includegraphics{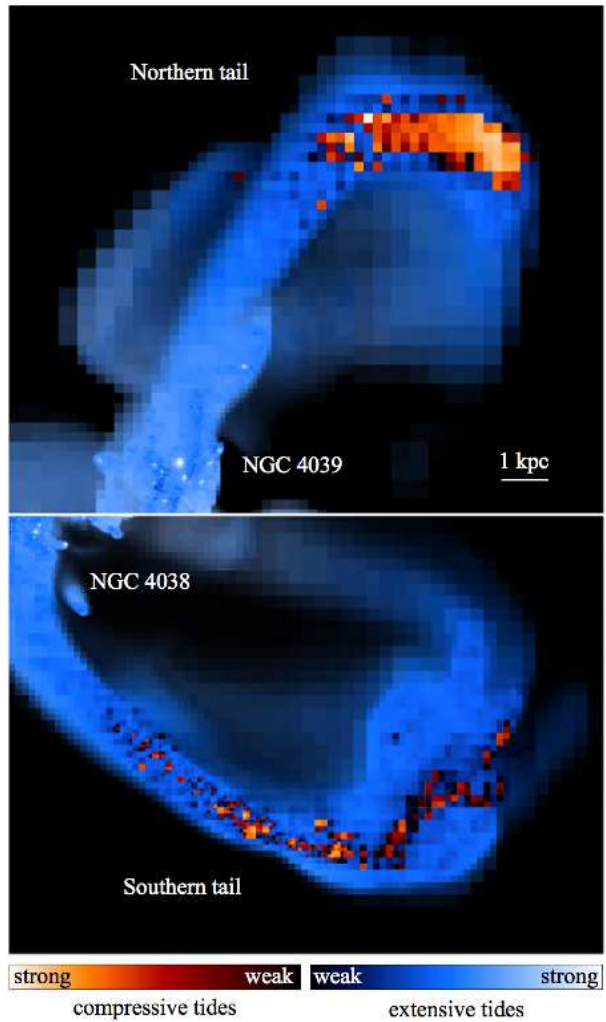}
\caption{Maps of the tidal field, at $t = 45 \Myr$ in the two tidal tails. The compressive tidal field is shown in orange-red, and the extensive field is in cyan-blue. The maps have been filtered to show only the dense regions, and hide the intergalactic medium.}
\label{fig:tdg_tides}
\end{figure}

Although neither the real Antennae nor our simulation contain any TDG, we can assume the key features of the scenarios accounting for their \emph{in situ} formations also apply to the smaller object we detect  \citep{Barnes1992, Elmegreen1993, Duc2004, Bournaud2004, Wetzstein2007}. Among them, \citet{Duc2004} proposed that gas accumulates into a massive and dense clump near the tip of the tail and then collapses to form a young stellar object. A slightly different scenario happens in our simulation: about $50 \Myr$ after the first pericenter passage, the bend in the Northern tail (visible at a later stage in \fig{morpho_gas}) is $9 \kpc$ away from the center of NGC~4039 and is moving away from the galaxy. The complex overlap of the potentials of the dark matter halos, the galaxy, and the tail itself creates a region of compressive tides over $\sim 3 \kpc \times 1 \kpc$ (\fig{tdg_tides}). Such tidal field helps the increase of the gas density and form a gas clump. The same process also takes place in the Southern tail, but the tidal field there is about 100 times weaker and spans several small regions, instead of a large one. The morphology of the tails (and thus the relative inclination of the discs at the pericenter passage which rules the efficiency of the spin-orbit coupling) seems to make most of the differences between the two tidal fields. In both tails, additional phenomena like low shear and the absence of kpc-scale streaming motions in this particular region further help the gathering of gas, but a seed $\sim 100$ times denser than the surrounding tail material only form in the Northern tail.

\begin{figure*}
\includegraphics{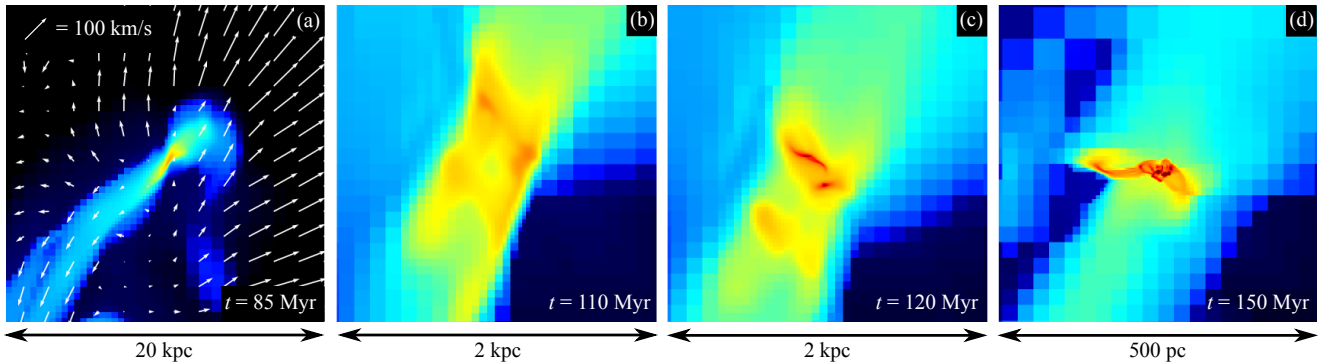}
\caption{Density maps of the gas in the Northern tidal tail, at several stages. In the left-most panel, the white arrows indicate the density-weighted velocity field averaged along the third, missing, dimension. The gas piles up close to the bend of the tail in the region, where the velocity field is the weakest, and where the tidal field was compressive a few $10 \Myr$ (\fig{tdg_tides}). The dense gas clumps shown in the right-most panel host the formation of a star cluster.}
\label{fig:tail_velocity}
\end{figure*}

Further away from the disc than our seed, the tip of the tail unfolds northward rapidly ($\sim 200 \mh 300 \kms$, because of its initial rotation speed from its progenitor disc). At the same time, the rest of the tail tends to fall back toward the galactic center, at comparable speeds but in the oposite direction, as visible in \fig{tail_velocity}.a. In between, the seed lies in a region of low velocity and accretes more and more gas ($\sim 50 \kms$), helped by an increasing self-gravity. Because of the geometry of the tail, the accretion becomes axisymetric. As a result, the tail narrows at this specific location, and the pinching transforms the seed into an elongated gas cloud. 

Two dense filaments ($\sim 1 \kpc$ long, $\sim 10^6 \Msun$) are formed on each side of the tail, both parallel to its axis (\fig{tail_velocity}.b). Such elongated over-densities could be associated to caustics on both edges of tidal tail \citep[see e.g.][on caustics propagating along tails]{Struck2012}. Then, the filaments collapse along their major axis under the effect of self-gravity. During the process, each structure keeps accreting surrounding gas and further collapses to finally form clouds (\fig{tail_velocity}.c). This fragmentation stage lasts about $20 \Myr$, i.e. of the order of the free-fall time of the initial filaments. At the end, one of the clumps does not show any sign of rotation (the western one, on the right-hand side in \fig{tail_velocity}.c) while the other spins at about $35 \kms$. Most probably, this difference originates in the slight asymmetry of the collapse of the second clump which thus accrete material with non-negligible angular momentum. Because of its fast rotation, the second cloud form small gaseous spiral arms. In the mean time, the first cloud, lacking the support from rotation, further collapses and gets dense enough to form stars ($t\approx 120 \Myr$).

\begin{figure}
\includegraphics{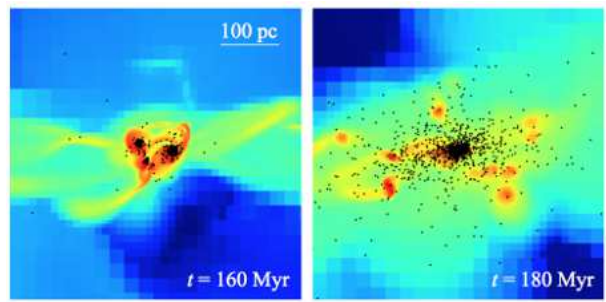}
\caption{Density maps ($500 \pc \times 500 \pc $) of the gas, and the position of the young stars formed in the Northern tidal tail, before the clusters merge (left), and after (right). The giant molecular association host about ten dense sub-clouds, while only one cluster can be detected.}
\label{fig:tdg_cluster}
\end{figure}

Interestingly enough, the two clouds, about $100 \pc$ from each other, with comparable masses ($\sim 10^7 \Msun$), formed at the same epoch and under comparable conditions (to some extent), show completely different star formation histories. The second, rotating cloud starts to form its own stars $15 \Myr$ after the first cloud. During this timelapse, a third cloud appears along a spiral arm of the second one. The three clumps interact with each other, keep accreting and form sub-clouds (\fig{tail_velocity}.d), following the idea of \citet{Bonnell2003} and similar to the formation scenario of W3 by \citet[see \sect{w3}]{Fellhauer2005} although the object formed here is 300 times more compact that W3. $150 \Myr$ after the creation of the tidal tail at the galactic pericenter passage, the three gas clumps host star formation, with no sign of any recent external trigger, contrarily to the clouds in the central regions of the galaxies. At the end of the simulation, the stellar objects have merged into a single cluster while the gas clouds are still separated, as shown in \fig{tdg_cluster}. The full star formation episode produces $3\times 10^7 \Msun$ over $75 \Myr$, with a peak at $1 \Msunyr$.

This stellar object appears to be much more compact than observed cases of TDGs \citep{Duc2014}. It has, in fact, the characteristics of a globular cluster (despite its strong rotation). We note that, at $t\approx 100 \Myr$, i.e. when the separation between the galaxies is maximum, this cluster started to fall back toward the galactic centers. Its high orbital energy (about twice larger than the other clusters in the simulation) suggests that, despite the energy loss it will experience when entering regions of high density close to the galactic center, it could maintain a large apocentric distance. The strong tidal field it would eventually experience when going close to the galactic center would significantly accelerate its mass-loss \citep{Renaud2013a}, and would slow down its rotation (by tidally stripping stars with high angular momentum, see \citealt{Varri2012} and references therein), making its properties more in line with the bulk of observed globular clusters. However, our simulation does not allow us to follow it for a sufficient time, nor to properly account for mass-loss effects to determine the precise fate of this object.

Note that only one star cluster is formed in our tidal tails, contrarily to the observations of several tens of clusters in tails of the Mice or the Tadpole galaxies by \citet{deGrijs2003}, and of ``beads on a string'' seen by \citet{Mullan2011}. \citet{Knierman2003} argued that tails could only host the formation of many small structures along them, or a massive TDG-like object, but not both. The presence in our simulated tails of a star cluster as massive as a TDG and the absence of any other cluster supports this conclusion.

It is difficult to foresee to what extend the scenario presented here is specific to our simulation, or could be extended to other setups, but this object demonstrates that globular clusters can be formed in mergers, even at large distances from the galactic center. Other globular clusters with comparable properties (size, mass) are also found in the central $\sim 10 \kpc$, indicating that the formation of a massive, dense stellar object in merger is not limited to the environment of tidal tails.

\section{Summary and conclusion}
\label{sec:summary}

We analyse the properties of star clusters in a hydrodynamical simulation of the Antennae galaxy merger at parsec-resolution, including star formation and stellar feedback (photoionisation, radiative pressure and supernovae). As shown in \citet{Renaud2014a}, the star formation rate becomes independent of the numerical resolution of the simulation at the scale of a few parsecs. Our main findings are as follows.
\begin{itemize}
\item Although each galactic encounter triggers a burst of star formation, the physics of the bursts strongly varies along the merger. In particular, the last collision enhances star formation mostly in the galactic center while earlier passages generate starburst over extended volumes.
\item More stars are formed during the ($\sim 100 \Myr$-long) separation of the galaxy progenitors, i.e. in between the encounters, than in each ($\sim 10 \Myr$-long) starburst episode of the merger sequence.
\item The different physical condition of each collision strongly impacts the formation of star clusters: most of the clusters are formed in the first galactic passages while the final coalescence does not trigger the formation of additional stellar systems at the parsec resolution.
\item The galactic interaction modifies the nature of the tidal field and the turbulence of the interstellar medium which translates into a starburst activity and the formation of clusters and dense stellar systems 30 times more massive than in isolated galaxies.
\item The typical star formation episode of star clusters lasts about $10 \Myr$. Some objects contain stars with a wider age spread, either because of the accretion of smaller clusters, or because of a secondary burst of star formation triggered by the kpc-scale dynamics of the merger.
\item A few objects formed have the properties of ultra-compact dwarf galaxies (several $10^7 \Msun$, $\gtrsim 10 \pc$). They share the formation mechanism of massive clusters. Their large half-mass radius originate either from their formation via a merger of gas clouds, or a violent expansion after flying through the galactic center. The hierarchical formation in sub-clouds being part of a giant complex, as proposed by \citet{Fellhauer2005} is not a unique way to form such objects.
\item One tidal tail shows the formation of a young massive cluster, too compact to be considered as a tidal dwarf galaxy. Its \emph{in situ} formation is favored by the kpc-scale dynamics of the system at a specific position along the tails. Compressive tides and the absence of shearing motions allow for the formation of a dense gaseous seed, its further accretion of surrounding material and finally star formation. This object is likely the progenitor of a globular cluster.
\end{itemize}

Despite the convergence of the star formation rate reached by our simulation, several important physical aspects are still missing. In particular, the collisional treatment of the gravitational force which rules the evolution of dense stellar systems through two-body relaxation is yet too numerically costly to be included at galactic scales. Thus, resolving the inner structure and internal dynamics of star clusters remains out-of-reach in present-day galaxy simulations. Pioneer steps have been taken to bridge this gap, e.g. by including time-dependent galactic tidal fields in $N$-body simulations of star clusters \citep{Renaud2013a, Rieder2013}, but a full coupling of the two scale is yet out-of-reach. The simulation presented here shows the way for a more realistic description of the initial conditions of the long-term evolution of star clusters.

As mentioned above, our simulation suggests that galaxy mergers could play a role in the formation of secondary populations of star in massive clusters. This topic deserves to be further explored. A more detailed description of stellar evolution and stellar feedback, together with a resolution of the sub-parsec structure of star clusters could provide interesting constraints on the effects of self-enrichment and the possible formation of multiple stellar populations. 

\section*{Acknowledgments}
We acknowledge fruitful and constructive comments from Nate Bastian, Sara Ellison, Bruce Elmegreen, Eric Emsellem, Mark Gieles, Richard De Grijs, Simon Karl, Doug Lin, Thorsten Naab, Rory Smith, Curt Struck and the anonymous referee. This work was supported by GENCI and PRACE resources (2013,2014-GEN2192 and pr86di allocations). FR and FB acknowledge support from the European Research Council through grant ERC-StG-257720.

\bibliographystyle{mn2e}

\begin{thebibliography}{}

\bibitem[\protect\citeauthoryear{{Alexander} \& {Gieles}}{{Alexander} \&
  {Gieles}}{2012}]{Alexander2012}
{Alexander} P.~E.~R.,  {Gieles} M.,  2012, \mnras, 422, 3415

\bibitem[\protect\citeauthoryear{{Barnes}}{{Barnes}}{1988}]{Barnes1988}
{Barnes} J.~E.,  1988, \apj, 331, 699

\bibitem[\protect\citeauthoryear{{Barnes}}{{Barnes}}{2004}]{Barnes2004}
{Barnes} J.~E.,  2004, \mnras, 350, 798

\bibitem[\protect\citeauthoryear{{Barnes} \& {Hernquist}}{{Barnes} \&
  {Hernquist}}{1992}]{Barnes1992}
{Barnes} J.~E.,  {Hernquist} L.,  1992, \araa, 30, 705

\bibitem[\protect\citeauthoryear{{Barnes} \& {Hernquist}}{{Barnes} \&
  {Hernquist}}{1996}]{Barnes1996}
{Barnes} J.~E.,  {Hernquist} L.,  1996, \apj, 471, 115

\bibitem[\protect\citeauthoryear{{Barnes} \& {Hernquist}}{{Barnes} \&
  {Hernquist}}{1991}]{Barnes1991}
{Barnes} J.~E.,  {Hernquist} L.~E.,  1991, \apjl, 370, L65

\bibitem[\protect\citeauthoryear{{Bastian}, {Cabrera-Ziri}, {Davies} \&
  {Larsen}}{{Bastian} et~al.}{2013}]{Bastian2013b}
{Bastian} N.,  {Cabrera-Ziri} I.,  {Davies} B.,    {Larsen} S.~S.,  2013,
  \mnras, 436, 2852

\bibitem[\protect\citeauthoryear{{Bastian}, {Schweizer}, {Goudfrooij}, {Larsen}
  \& {Kissler-Patig}}{{Bastian} et~al.}{2013}]{Bastian2013}
{Bastian} N.,  {Schweizer} F.,  {Goudfrooij} P.,  {Larsen} S.~S.,
  {Kissler-Patig} M.,  2013, \mnras, 431, 1252

\bibitem[\protect\citeauthoryear{{Bastian}, {Trancho}, {Konstantopoulos} \&
  {Miller}}{{Bastian} et~al.}{2009}]{Bastian2009}
{Bastian} N.,  {Trancho} G.,  {Konstantopoulos} I.~S.,    {Miller} B.~W.,
  2009, \apj, 701, 607

\bibitem[\protect\citeauthoryear{{Bekki} \& {Couch}}{{Bekki} \&
  {Couch}}{2001}]{Bekki2001}
{Bekki} K.,  {Couch} W.~J.,  2001, \apjl, 557, L19

\bibitem[\protect\citeauthoryear{{Bekki}, {Forbes}, {Beasley} \&
  {Couch}}{{Bekki} et~al.}{2002}]{Bekki2002}
{Bekki} K.,  {Forbes} D.~A.,  {Beasley} M.~A.,    {Couch} W.~J.,  2002, \mnras,
  335, 1176

\bibitem[\protect\citeauthoryear{{Berentzen}, {Athanassoula}, {Heller} \&
  {Fricke}}{{Berentzen} et~al.}{2004}]{Berentzen2004}
{Berentzen} I.,  {Athanassoula} E.,  {Heller} C.~H.,    {Fricke} K.~J.,  2004,
  \mnras, 347, 220

\bibitem[\protect\citeauthoryear{{Boily} \& {Kroupa}}{{Boily} \&
  {Kroupa}}{2003a}]{Boily2003a}
{Boily} C.~M.,  {Kroupa} P.,  2003a, \mnras, 338, 665

\bibitem[\protect\citeauthoryear{{Boily} \& {Kroupa}}{{Boily} \&
  {Kroupa}}{2003b}]{Boily2003b}
{Boily} C.~M.,  {Kroupa} P.,  2003b, \mnras, 338, 673

\bibitem[\protect\citeauthoryear{{Bonnell}, {Bate} \& {Vine}}{{Bonnell}
  et~al.}{2003}]{Bonnell2003}
{Bonnell} I.~A.,  {Bate} M.~R.,    {Vine} S.~G.,  2003, \mnras, 343, 413

\bibitem[\protect\citeauthoryear{{Bournaud}, {Duc} \& {Emsellem}}{{Bournaud}
  et~al.}{2008}]{Bournaud2008}
{Bournaud} F.,  {Duc} P.,    {Emsellem} E.,  2008, \mnras, 389, L8

\bibitem[\protect\citeauthoryear{{Bournaud}, {Duc}, {Amram}, {Combes} \&
  {Gach}}{{Bournaud} et~al.}{2004}]{Bournaud2004}
{Bournaud} F.,  {Duc} P.-A.,  {Amram} P.,  {Combes} F.,    {Gach} J.-L.,  2004,
  \aap, 425, 813

\bibitem[\protect\citeauthoryear{{Bournaud}, {Elmegreen}, {Teyssier}, {Block}
  \& {Puerari}}{{Bournaud} et~al.}{2010}]{Bournaud2010b}
{Bournaud} F.,  {Elmegreen} B.~G.,  {Teyssier} R.,  {Block} D.~L.,    {Puerari}
  I.,  2010, \mnras, 409, 1088

\bibitem[\protect\citeauthoryear{{Bournaud}, {Perret}, {Renaud}, {Dekel},
  {Elmegreen}, {Elmegreen}, {Teyssier}, {Amram} \& {et al.}}{{Bournaud}
  et~al.}{2014}]{Bournaud2014}
{Bournaud} F.,  {Perret} V.,  {Renaud} F.,  {Dekel} A.,  {Elmegreen} B.~G.,
  {Elmegreen} D.~M.,  {Teyssier} R.,  {Amram} P.,    {et al.} 2014, \apj, 780,
  57

\bibitem[\protect\citeauthoryear{{Boutloukos} \& {Lamers}}{{Boutloukos} \&
  {Lamers}}{2003}]{Boutloukos2003}
{Boutloukos} S.~G.,  {Lamers} H.~J.~G.~L.~M.,  2003, \mnras, 338, 717

\bibitem[\protect\citeauthoryear{{Burkert}}{{Burkert}}{1995}]{Burkert1995}
{Burkert} A.,  1995, \apjl, 447, L25

\bibitem[\protect\citeauthoryear{{Cabrera-Ziri}, {Bastian}, {Davies}, {Magris},
  {Bruzual} \& {Schweizer}}{{Cabrera-Ziri} et~al.}{2014}]{Cabrera2014}
{Cabrera-Ziri} I.,  {Bastian} N.,  {Davies} B.,  {Magris} G.,  {Bruzual} G.,
  {Schweizer} F.,  2014, \mnras, 441, 2754

\bibitem[\protect\citeauthoryear{{Caso}, {Bassino}, {Richtler}, {Calder{\'o}n}
  \& {Smith Castelli}}{{Caso} et~al.}{2014}]{Caso2014}
{Caso} J.~P.,  {Bassino} L.~P.,  {Richtler} T.,  {Calder{\'o}n} J.~P.,
  {Smith Castelli} A.~V.,  2014, \mnras, 442, 891

\bibitem[\protect\citeauthoryear{{C{\^o}t{\'e}}, {Piatek}, {Ferrarese},
  {Jord{\'a}n}, {Merritt}, {Peng}, {Ha{\c s}egan}, {Blakeslee}, {Mei}, {West},
  {Milosavljevi{\'c}} \& {Tonry}}{{C{\^o}t{\'e}} et~al.}{2006}]{Cote2006}
{C{\^o}t{\'e}} P.,  {Piatek} S.,  {Ferrarese} L.,  {Jord{\'a}n} A.,  {Merritt}
  D.,  {Peng} E.~W.,  {Ha{\c s}egan} M.,  {Blakeslee} J.~P.,  {Mei} S.,  {West}
  M.~J.,  {Milosavljevi{\'c}} M.,    {Tonry} J.~L.,  2006, \apjs, 165, 57

\bibitem[\protect\citeauthoryear{{Courty} \& {Alimi}}{{Courty} \&
  {Alimi}}{2004}]{Courty2004}
{Courty} S.,  {Alimi} J.~M.,  2004, \aap, 416, 875

\bibitem[\protect\citeauthoryear{{Cox}, {Jonsson}, {Primack} \&
  {Somerville}}{{Cox} et~al.}{2006}]{Cox2006}
{Cox} T.~J.,  {Jonsson} P.,  {Primack} J.~R.,    {Somerville} R.~S.,  2006,
  \mnras, 373, 1013

\bibitem[\protect\citeauthoryear{{Cox}, {Jonsson}, {Somerville}, {Primack} \&
  {Dekel}}{{Cox} et~al.}{2008}]{Cox2008}
{Cox} T.~J.,  {Jonsson} P.,  {Somerville} R.~S.,  {Primack} J.~R.,    {Dekel}
  A.,  2008, \mnras, 384, 386

\bibitem[\protect\citeauthoryear{{de Grijs}, {Lee}, {Clemencia Mora Herrera},
  {Fritze-v.~Alvensleben} \& {Anders}}{{de Grijs} et~al.}{2003}]{deGrijs2003}
{de Grijs} R.,  {Lee} J.~T.,  {Clemencia Mora Herrera} M.,
  {Fritze-v.~Alvensleben} U.,    {Anders} P.,  2003, \na, 8, 155

\bibitem[\protect\citeauthoryear{{Di Matteo}, {Combes}, {Melchior} \&
  {Semelin}}{{Di Matteo} et~al.}{2007}]{DiMatteo2007}
{Di Matteo} P.,  {Combes} F.,  {Melchior} A.,    {Semelin} B.,  2007, \aap,
  468, 61

\bibitem[\protect\citeauthoryear{{Dubois} \& {Teyssier}}{{Dubois} \&
  {Teyssier}}{2008}]{Dubois2008}
{Dubois} Y.,  {Teyssier} R.,  2008, \aap, 477, 79

\bibitem[\protect\citeauthoryear{{Duc}, {Bournaud} \& {Masset}}{{Duc}
  et~al.}{2004}]{Duc2004}
{Duc} P.-A.,  {Bournaud} F.,    {Masset} F.,  2004, \aap, 427, 803

\bibitem[\protect\citeauthoryear{{Duc}, {Paudel}, {McDermid}, {Cuillandre},
  {Serra}, {Bournaud}, {Cappellari} \& {Emsellem}}{{Duc}
  et~al.}{2014}]{Duc2014}
{Duc} P.-A.,  {Paudel} S.,  {McDermid} R.~M.,  {Cuillandre} J.-C.,  {Serra} P.,
   {Bournaud} F.,  {Cappellari} M.,    {Emsellem} E.,  2014, \mnras, 440, 1458

\bibitem[\protect\citeauthoryear{{Eisenstein} \& {Hut}}{{Eisenstein} \&
  {Hut}}{1998}]{Eisenstein1998}
{Eisenstein} D.~J.,  {Hut} P.,  1998, \apj, 498, 137

\bibitem[\protect\citeauthoryear{{Elmegreen} \& {Elmegreen}}{{Elmegreen} \&
  {Elmegreen}}{2001}]{Elmegreen2001c}
{Elmegreen} B.~G.,  {Elmegreen} D.~M.,  2001, \aj, 121, 1507

\bibitem[\protect\citeauthoryear{{Elmegreen}, {Kaufman} \&
  {Thomasson}}{{Elmegreen} et~al.}{1993}]{Elmegreen1993}
{Elmegreen} B.~G.,  {Kaufman} M.,    {Thomasson} M.,  1993, \apj, 412, 90

\bibitem[\protect\citeauthoryear{{Elmegreen}, {Elmegreen}, {Adamo}, {Aloisi},
  {Andrews}, {Annibali}, {Bright}, {Calzetti} \& {et al.}}{{Elmegreen}
  et~al.}{2014}]{Elmegreen2014}
{Elmegreen} D.~M.,  {Elmegreen} B.~G.,  {Adamo} A.,  {Aloisi} A.,  {Andrews}
  J.,  {Annibali} F.,  {Bright} S.~N.,  {Calzetti} D.,    {et al.} 2014, \apjl,
  787, L15

\bibitem[\protect\citeauthoryear{{Fall}, {Chandar} \& {Whitmore}}{{Fall}
  et~al.}{2005}]{Fall2005}
{Fall} S.~M.,  {Chandar} R.,    {Whitmore} B.~C.,  2005, \apjl, 631, L133

\bibitem[\protect\citeauthoryear{{Fellhauer} \& {Kroupa}}{{Fellhauer} \&
  {Kroupa}}{2005}]{Fellhauer2005}
{Fellhauer} M.,  {Kroupa} P.,  2005, \mnras, 359, 223

\bibitem[\protect\citeauthoryear{{Fujimoto}, {Tasker}, {Wakayama} \&
  {Habe}}{{Fujimoto} et~al.}{2014}]{Fujimoto2014}
{Fujimoto} Y.,  {Tasker} E.~J.,  {Wakayama} M.,    {Habe} A.,  2014, \mnras,
  439, 936

\bibitem[\protect\citeauthoryear{{Fukui}, {Ohama}, {Hanaoka}, {Furukawa},
  {Torii}, {Dawson}, {Mizuno}, {Hasegawa} \& {et al.}}{{Fukui}
  et~al.}{2014}]{Fukui2014}
{Fukui} Y.,  {Ohama} A.,  {Hanaoka} N.,  {Furukawa} N.,  {Torii} K.,  {Dawson}
  J.~R.,  {Mizuno} N.,  {Hasegawa} K.,    {et al.} 2014, \apj, 780, 36

\bibitem[\protect\citeauthoryear{{Furukawa}, {Dawson}, {Ohama}, {Kawamura},
  {Mizuno}, {Onishi} \& {Fukui}}{{Furukawa} et~al.}{2009}]{Furukawa2009}
{Furukawa} N.,  {Dawson} J.~R.,  {Ohama} A.,  {Kawamura} A.,  {Mizuno} N.,
  {Onishi} T.,    {Fukui} Y.,  2009, \apjl, 696, L115

\bibitem[\protect\citeauthoryear{{Gieles}, {Heggie} \& {Zhao}}{{Gieles}
  et~al.}{2011}]{Gieles2011b}
{Gieles} M.,  {Heggie} D.~C.,    {Zhao} H.,  2011, \mnras, 413, 2509

\bibitem[\protect\citeauthoryear{{Gieles}, {Portegies Zwart}, {Baumgardt},
  {Athanassoula}, {Lamers}, {Sipior} \& {Leenaarts}}{{Gieles}
  et~al.}{2006}]{Gieles2006c}
{Gieles} M.,  {Portegies Zwart} S.~F.,  {Baumgardt} H.,  {Athanassoula} E.,
  {Lamers} H.~J.~G.~L.~M.,  {Sipior} M.,    {Leenaarts} J.,  2006, \mnras, 371,
  793

\bibitem[\protect\citeauthoryear{{Hernquist}}{{Hernquist}}{1990}]{Hernquist1990}
{Hernquist} L.,  1990, \apj, 356, 359

\bibitem[\protect\citeauthoryear{{Herrera}, {Boulanger} \&
  {Nesvadba}}{{Herrera} et~al.}{2011}]{Herrera2011}
{Herrera} C.~N.,  {Boulanger} F.,    {Nesvadba} N.~P.~H.,  2011, \aap, 534,
  A138

\bibitem[\protect\citeauthoryear{{Hopkins}, {Cox}, {Younger} \&
  {Hernquist}}{{Hopkins} et~al.}{2009}]{Hopkins2009}
{Hopkins} P.~F.,  {Cox} T.~J.,  {Younger} J.~D.,    {Hernquist} L.,  2009,
  \apj, 691, 1168

\bibitem[\protect\citeauthoryear{{Jog} \& {Solomon}}{{Jog} \&
  {Solomon}}{1992}]{Jog1992}
{Jog} C.~J.,  {Solomon} P.~M.,  1992, \apj, 387, 152

\bibitem[\protect\citeauthoryear{{Jord{\'a}n}, {Peng}, {Blakeslee},
  {C{\^o}t{\'e}}, {Eyheramendy}, {Ferrarese}, {Mei}, {Tonry} \&
  {West}}{{Jord{\'a}n} et~al.}{2009}]{Jordan2009}
{Jord{\'a}n} A.,  {Peng} E.~W.,  {Blakeslee} J.~P.,  {C{\^o}t{\'e}} P.,
  {Eyheramendy} S.,  {Ferrarese} L.,  {Mei} S.,  {Tonry} J.~L.,    {West}
  M.~J.,  2009, \apjs, 180, 54

\bibitem[\protect\citeauthoryear{{Karl}, {Fall} \& {Naab}}{{Karl}
  et~al.}{2011}]{Karl2011}
{Karl} S.~J.,  {Fall} S.~M.,    {Naab} T.,  2011, \apj, 734, 11

\bibitem[\protect\citeauthoryear{{Karl}, {Lunttila}, {Naab}, {Johansson},
  {Klaas} \& {Juvela}}{{Karl} et~al.}{2013}]{Karl2013}
{Karl} S.~J.,  {Lunttila} T.,  {Naab} T.,  {Johansson} P.~H.,  {Klaas} U.,
  {Juvela} M.,  2013, \mnras, 434, 696

\bibitem[\protect\citeauthoryear{{Karl}, {Naab}, {Johansson}, {Kotarba},
  {Boily}, {Renaud} \& {Theis}}{{Karl} et~al.}{2010}]{Karl2010}
{Karl} S.~J.,  {Naab} T.,  {Johansson} P.~H.,  {Kotarba} H.,  {Boily} C.~M.,
  {Renaud} F.,    {Theis} C.,  2010, \apjl, 715, L88

\bibitem[\protect\citeauthoryear{{Knierman}, {Gallagher}, {Charlton},
  {Hunsberger}, {Whitmore}, {Kundu}, {Hibbard} \& {Zaritsky}}{{Knierman}
  et~al.}{2003}]{Knierman2003}
{Knierman} K.~A.,  {Gallagher} S.~C.,  {Charlton} J.~C.,  {Hunsberger} S.~D.,
  {Whitmore} B.,  {Kundu} A.,  {Hibbard} J.~E.,    {Zaritsky} D.,  2003, \aj,
  126, 1227

\bibitem[\protect\citeauthoryear{{Kraljic}, {Renaud}, {Bournaud}, {Combes},
  {Elmegreen}, {Emsellem} \& {Teyssier}}{{Kraljic} et~al.}{2014}]{Kraljic2014}
{Kraljic} K.,  {Renaud} F.,  {Bournaud} F.,  {Combes} F.,  {Elmegreen} B.,
  {Emsellem} E.,    {Teyssier} R.,  2014, \apj, 784, 112

\bibitem[\protect\citeauthoryear{{Larsen}}{{Larsen}}{2009}]{Larsen2009}
{Larsen} S.~S.,  2009, \aap, 494, 539

\bibitem[\protect\citeauthoryear{{Li}, {Mac Low} \& {Klessen}}{{Li}
  et~al.}{2004}]{Li2004}
{Li} Y.,  {Mac Low} M.-M.,    {Klessen} R.~S.,  2004, \apjl, 614, L29

\bibitem[\protect\citeauthoryear{{Ma{\'{\i}}z-Apell{\'a}niz}}{{Ma{\'{\i}}z-Apell{\'a}niz}}{2001}]{Maiz2001}
{Ma{\'{\i}}z-Apell{\'a}niz} J.,  2001, \apj, 563, 151

\bibitem[\protect\citeauthoryear{{Maraston}, {Bastian}, {Saglia},
  {Kissler-Patig}, {Schweizer} \& {Goudfrooij}}{{Maraston}
  et~al.}{2004}]{Maraston2004}
{Maraston} C.,  {Bastian} N.,  {Saglia} R.~P.,  {Kissler-Patig} M.,
  {Schweizer} F.,    {Goudfrooij} P.,  2004, \aap, 416, 467

\bibitem[\protect\citeauthoryear{{Matsui}, {Saitoh}, {Makino}, {Wada},
  {Tomisaka}, {Kokubo}, {Daisaka}, {Okamoto} \& {Yoshida}}{{Matsui}
  et~al.}{2012}]{Matsui2012}
{Matsui} H.,  {Saitoh} T.~R.,  {Makino} J.,  {Wada} K.,  {Tomisaka} K.,
  {Kokubo} E.,  {Daisaka} H.,  {Okamoto} T.,    {Yoshida} N.,  2012, \apj, 746,
  26

\bibitem[\protect\citeauthoryear{{McLaughlin} \& {van der Marel}}{{McLaughlin}
  \& {van der Marel}}{2005}]{McLaughlin2005}
{McLaughlin} D.~E.,  {van der Marel} R.~P.,  2005, \apjs, 161, 304

\bibitem[\protect\citeauthoryear{{Mengel}, {Lehnert}, {Thatte} \&
  {Genzel}}{{Mengel} et~al.}{2005}]{Mengel2005}
{Mengel} S.,  {Lehnert} M.~D.,  {Thatte} N.,    {Genzel} R.,  2005, \aap, 443,
  41

\bibitem[\protect\citeauthoryear{{Mengel}, {Lehnert}, {Thatte}, {Vacca},
  {Whitmore} \& {Chandar}}{{Mengel} et~al.}{2008}]{Mengel2008}
{Mengel} S.,  {Lehnert} M.~D.,  {Thatte} N.~A.,  {Vacca} W.~D.,  {Whitmore} B.,
     {Chandar} R.,  2008, \aap, 489, 1091

\bibitem[\protect\citeauthoryear{{Mieske}, {Hilker}, {Jord{\'a}n}, {Infante},
  {Kissler-Patig}, {Rejkuba}, {Richtler}, {C{\^o}t{\'e}}, {Baumgardt}, {West},
  {Ferrarese} \& {Peng}}{{Mieske} et~al.}{2008}]{Mieske2008}
{Mieske} S.,  {Hilker} M.,  {Jord{\'a}n} A.,  {Infante} L.,  {Kissler-Patig}
  M.,  {Rejkuba} M.,  {Richtler} T.,  {C{\^o}t{\'e}} P.,  {Baumgardt} H.,
  {West} M.~J.,  {Ferrarese} L.,    {Peng} E.~W.,  2008, \aap, 487, 921

\bibitem[\protect\citeauthoryear{{Mieske}, {Hilker} \& {Misgeld}}{{Mieske}
  et~al.}{2012}]{Mieske2012}
{Mieske} S.,  {Hilker} M.,    {Misgeld} I.,  2012, \aap, 537, A3

\bibitem[\protect\citeauthoryear{{Mihos}, {Bothun} \& {Richstone}}{{Mihos}
  et~al.}{1993}]{Mihos1993}
{Mihos} J.~C.,  {Bothun} G.~D.,    {Richstone} D.~O.,  1993, \apj, 418, 82

\bibitem[\protect\citeauthoryear{{Mihos} \& {Hernquist}}{{Mihos} \&
  {Hernquist}}{1996}]{Mihos1996}
{Mihos} J.~C.,  {Hernquist} L.,  1996, \apj, 464, 641

\bibitem[\protect\citeauthoryear{{Misgeld} \& {Hilker}}{{Misgeld} \&
  {Hilker}}{2011}]{Misgeld2011}
{Misgeld} I.,  {Hilker} M.,  2011, \mnras, 414, 3699

\bibitem[\protect\citeauthoryear{{Moreno}, {Bluck}, {Ellison}, {Patton},
  {Torrey} \& {Moster}}{{Moreno} et~al.}{2013}]{Moreno2013}
{Moreno} J.,  {Bluck} A.~F.~L.,  {Ellison} S.~L.,  {Patton} D.~R.,  {Torrey}
  P.,    {Moster} B.~P.,  2013, \mnras, 436, 1765

\bibitem[\protect\citeauthoryear{{Motte}, {Nguyen Luong}, {Schneider},
  {Heitsch}, {Glover}, {Carlhoff}, {Hill}, {Bontemps}, {Schilke}, {Louvet},
  {Hennemann}, {Didelon} \& {Beuther}}{{Motte} et~al.}{2014}]{Motte2014}
{Motte} F.,  {Nguyen Luong} Q.,  {Schneider} N.,  {Heitsch} F.,  {Glover} S.,
  {Carlhoff} P.,  {Hill} T.,  {Bontemps} S.,  {Schilke} P.,  {Louvet} F.,
  {Hennemann} M.,  {Didelon} P.,    {Beuther} H.,  2014, ArXiv e-prints

\bibitem[\protect\citeauthoryear{{Mullan}, {Konstantopoulos}, {Kepley}, {Lee},
  {Charlton}, {Knierman}, {Bastian}, {Chandar} \& {et al.}}{{Mullan}
  et~al.}{2011}]{Mullan2011}
{Mullan} B.,  {Konstantopoulos} I.~S.,  {Kepley} A.~A.,  {Lee} K.~H.,
  {Charlton} J.~C.,  {Knierman} K.,  {Bastian} N.,  {Chandar} R.,    {et al.}
  2011, \apj, 731, 93

\bibitem[\protect\citeauthoryear{{Naab} \& {Burkert}}{{Naab} \&
  {Burkert}}{2003}]{Naab2003}
{Naab} T.,  {Burkert} A.,  2003, \apj, 597, 893

\bibitem[\protect\citeauthoryear{{Norris}, {Kannappan}, {Forbes}, {Romanowsky},
  {Brodie}, {Faifer}, {Huxor}, {Maraston} \& {et al.}}{{Norris}
  et~al.}{2014}]{Norris2014}
{Norris} M.~A.,  {Kannappan} S.~J.,  {Forbes} D.~A.,  {Romanowsky} A.~J.,
  {Brodie} J.~P.,  {Faifer} F.~R.,  {Huxor} A.,  {Maraston} C.,    {et al.}
  2014, \mnras, 443, 1151

\bibitem[\protect\citeauthoryear{{Perret}, {Renaud}, {Epinat}, {Amram},
  {Bournaud}, {Contini}, {Teyssier} \& {Lambert}}{{Perret}
  et~al.}{2014}]{Perret2014}
{Perret} V.,  {Renaud} F.,  {Epinat} B.,  {Amram} P.,  {Bournaud} F.,
  {Contini} T.,  {Teyssier} R.,    {Lambert} J.-C.,  2014, \aap, 562, A1

\bibitem[\protect\citeauthoryear{{Peterson}, {Struck}, {Smith} \&
  {Hancock}}{{Peterson} et~al.}{2009}]{Peterson2009}
{Peterson} B.~W.,  {Struck} C.,  {Smith} B.~J.,    {Hancock} M.,  2009, \mnras,
  400, 1208

\bibitem[\protect\citeauthoryear{{Powell}, {Bournaud}, {Chapon} \&
  {Teyssier}}{{Powell} et~al.}{2013}]{Powell2013}
{Powell} L.~C.,  {Bournaud} F.,  {Chapon} D.,    {Teyssier} R.,  2013, \mnras,
  434, 1028

\bibitem[\protect\citeauthoryear{{Privon}, {Barnes}, {Evans}, {Hibbard}, {Yun},
  {Mazzarella}, {Armus} \& {Surace}}{{Privon} et~al.}{2013}]{Privon2013}
{Privon} G.~C.,  {Barnes} J.~E.,  {Evans} A.~S.,  {Hibbard} J.~E.,  {Yun}
  M.~S.,  {Mazzarella} J.~M.,  {Armus} L.,    {Surace} J.,  2013, \apj, 771,
  120

\bibitem[\protect\citeauthoryear{{Renaud}, {Boily}, {Fleck}, {Naab} \&
  {Theis}}{{Renaud} et~al.}{2008}]{Renaud2008}
{Renaud} F.,  {Boily} C.~M.,  {Fleck} J.-J.,  {Naab} T.,    {Theis} C.,  2008,
  \mnras, 391, L98

\bibitem[\protect\citeauthoryear{{Renaud}, {Boily}, {Naab} \& {Theis}}{{Renaud}
  et~al.}{2009}]{Renaud2009}
{Renaud} F.,  {Boily} C.~M.,  {Naab} T.,    {Theis} C.,  2009, \apj, 706, 67

\bibitem[\protect\citeauthoryear{{Renaud}, {Bournaud}, {Emsellem}, {Elmegreen},
  {Teyssier}, {Alves}, {Chapon}, {Combes} \& {et al.}}{{Renaud}
  et~al.}{2013}]{Renaud2013b}
{Renaud} F.,  {Bournaud} F.,  {Emsellem} E.,  {Elmegreen} B.,  {Teyssier} R.,
  {Alves} J.,  {Chapon} D.,  {Combes} F.,    {et al.} 2013, \mnras, 436, 1836

\bibitem[\protect\citeauthoryear{{Renaud}, {Bournaud}, {Kraljic} \&
  {Duc}}{{Renaud} et~al.}{2014}]{Renaud2014a}
{Renaud} F.,  {Bournaud} F.,  {Kraljic} K.,    {Duc} P.-A.,  2014, \mnras, 442,
  L33

\bibitem[\protect\citeauthoryear{{Renaud} \& {Gieles}}{{Renaud} \&
  {Gieles}}{2013}]{Renaud2013a}
{Renaud} F.,  {Gieles} M.,  2013, \mnras, 431, L83

\bibitem[\protect\citeauthoryear{{Rieder}, {Ishiyama}, {Langelaan}, {Makino},
  {McMillan} \& {Portegies Zwart}}{{Rieder} et~al.}{2013}]{Rieder2013}
{Rieder} S.,  {Ishiyama} T.,  {Langelaan} P.,  {Makino} J.,  {McMillan}
  S.~L.~W.,    {Portegies Zwart} S.,  2013, \mnras, 436, 3695

\bibitem[\protect\citeauthoryear{{Robertson} \& {Kravtsov}}{{Robertson} \&
  {Kravtsov}}{2008}]{Robertson2008}
{Robertson} B.~E.,  {Kravtsov} A.~V.,  2008, \apj, 680, 1083

\bibitem[\protect\citeauthoryear{{Saitoh}, {Daisaka}, {Kokubo}, {Makino},
  {Okamoto}, {Tomisaka}, {Wada} \& {Yoshida}}{{Saitoh}
  et~al.}{2009}]{Saitoh2009}
{Saitoh} T.~R.,  {Daisaka} H.,  {Kokubo} E.,  {Makino} J.,  {Okamoto} T.,
  {Tomisaka} K.,  {Wada} K.,    {Yoshida} N.,  2009, \pasj, 61, 481

\bibitem[\protect\citeauthoryear{{Saviane}, {Momany}, {da Costa}, {Rich} \&
  {Hibbard}}{{Saviane} et~al.}{2008}]{Saviane2008}
{Saviane} I.,  {Momany} Y.,  {da Costa} G.~S.,  {Rich} R.~M.,    {Hibbard}
  J.~E.,  2008, \apj, 678, 179

\bibitem[\protect\citeauthoryear{{Schweizer}, {Burns}, {Madore}, {Mager},
  {Phillips}, {Freedman}, {Boldt}, {Contreras}, {Folatelli}, {Gonz{\'a}lez},
  {Hamuy}, {Krzeminski}, {Morrell}, {Persson}, {Roth} \&
  {Stritzinger}}{{Schweizer} et~al.}{2008}]{Schweizer2008}
{Schweizer} F.,  {Burns} C.~R.,  {Madore} B.~F.,  {Mager} V.~A.,  {Phillips}
  M.~M.,  {Freedman} W.~L.,  {Boldt} L.,  {Contreras} C.,  {Folatelli} G.,
  {Gonz{\'a}lez} S.,  {Hamuy} M.,  {Krzeminski} W.,  {Morrell} N.~I.,
  {Persson} S.~E.,  {Roth} M.~R.,    {Stritzinger} M.~D.,  2008, \aj, 136, 1482

\bibitem[\protect\citeauthoryear{{Smith}, {Soria}, {Struck}, {Giroux}, {Swartz}
  \& {Yukita}}{{Smith} et~al.}{2014}]{Smith2014}
{Smith} B.~J.,  {Soria} R.,  {Struck} C.,  {Giroux} M.~L.,  {Swartz} D.~A.,
  {Yukita} M.,  2014, \aj, 147, 60

\bibitem[\protect\citeauthoryear{{Steinmetz} \& {White}}{{Steinmetz} \&
  {White}}{1997}]{Steinmetz1997}
{Steinmetz} M.,  {White} S.~D.~M.,  1997, \mnras, 288, 545

\bibitem[\protect\citeauthoryear{{Struck} \& {Smith}}{{Struck} \&
  {Smith}}{2003}]{Struck2003}
{Struck} C.,  {Smith} B.~J.,  2003, \apj, 589, 157

\bibitem[\protect\citeauthoryear{{Struck} \& {Smith}}{{Struck} \&
  {Smith}}{2012}]{Struck2012}
{Struck} C.,  {Smith} B.~J.,  2012, \mnras, 422, 2444

\bibitem[\protect\citeauthoryear{{Teyssier}}{{Teyssier}}{2002}]{Teyssier2002}
{Teyssier} R.,  2002, \aap, 385, 337

\bibitem[\protect\citeauthoryear{{Teyssier}, {Chapon} \& {Bournaud}}{{Teyssier}
  et~al.}{2010}]{Teyssier2010}
{Teyssier} R.,  {Chapon} D.,    {Bournaud} F.,  2010, \apjl, 720, L149

\bibitem[\protect\citeauthoryear{{Teyssier}, {Pontzen}, {Dubois} \&
  {Read}}{{Teyssier} et~al.}{2013}]{Teyssier2013}
{Teyssier} R.,  {Pontzen} A.,  {Dubois} Y.,    {Read} J.~I.,  2013, \mnras,
  429, 3068

\bibitem[\protect\citeauthoryear{{Toomre} \& {Toomre}}{{Toomre} \&
  {Toomre}}{1972}]{Toomre1972}
{Toomre} A.,  {Toomre} J.,  1972, \apj, 178, 623

\bibitem[\protect\citeauthoryear{{Truelove}, {Klein}, {McKee}, {Holliman} II,
  {Howell} \& {Greenough}}{{Truelove} et~al.}{1997}]{Truelove1997}
{Truelove} J.~K.,  {Klein} R.~I.,  {McKee} C.~F.,  {Holliman} II J.~H.,
  {Howell} L.~H.,    {Greenough} J.~A.,  1997, \apjl, 489, L179

\bibitem[\protect\citeauthoryear{{Varri} \& {Bertin}}{{Varri} \&
  {Bertin}}{2012}]{Varri2012}
{Varri} A.~L.,  {Bertin} G.,  2012, \aap, 540, A94

\bibitem[\protect\citeauthoryear{{Wetzstein}, {Naab} \& {Burkert}}{{Wetzstein}
  et~al.}{2007}]{Wetzstein2007}
{Wetzstein} M.,  {Naab} T.,    {Burkert} A.,  2007, \mnras, 375, 805

\bibitem[\protect\citeauthoryear{{Whitmore}, {Chandar} \& {Fall}}{{Whitmore}
  et~al.}{2007}]{Whitmore2007}
{Whitmore} B.~C.,  {Chandar} R.,    {Fall} S.~M.,  2007, \aj, 133, 1067

\bibitem[\protect\citeauthoryear{{Whitmore}, {Chandar}, {Schweizer},
  {Rothberg}, {Leitherer}, {Rieke}, {Rieke}, {Blair}, {Mengel} \&
  {Alonso-Herrero}}{{Whitmore} et~al.}{2010}]{Whitmore2010}
{Whitmore} B.~C.,  {Chandar} R.,  {Schweizer} F.,  {Rothberg} B.,  {Leitherer}
  C.,  {Rieke} M.,  {Rieke} G.,  {Blair} W.~P.,  {Mengel} S.,
  {Alonso-Herrero} A.,  2010, \aj, 140, 75

\bibitem[\protect\citeauthoryear{{Whitmore} \& {Schweizer}}{{Whitmore} \&
  {Schweizer}}{1995}]{Whitmore1995}
{Whitmore} B.~C.,  {Schweizer} F.,  1995, \aj, 109, 960

\bibitem[\protect\citeauthoryear{{Whitmore}, {Zhang}, {Leitherer}, {Fall},
  {Schweizer} \& {Miller}}{{Whitmore} et~al.}{1999}]{Whitmore1999a}
{Whitmore} B.~C.,  {Zhang} Q.,  {Leitherer} C.,  {Fall} S.~M.,  {Schweizer} F.,
     {Miller} B.~W.,  1999, \aj, 118, 1551

\bibitem[\protect\citeauthoryear{{Wolf}, {Bell}, {McIntosh}, {Rix}, {Barden},
  {Beckwith}, {Borch}, {Caldwell}, {H{\"a}ussler}, {Heymans}, {Jahnke},
  {Jogee}, {Meisenheimer}, {Peng}, {S{\'a}nchez}, {Somerville} \&
  {Wisotzki}}{{Wolf} et~al.}{2005}]{Wolf2005}
{Wolf} C.,  {Bell} E.~F.,  {McIntosh} D.~H.,  {Rix} H.-W.,  {Barden} M.,
  {Beckwith} S.~V.~W.,  {Borch} A.,  {Caldwell} J.~A.~R.,  {H{\"a}ussler} B.,
  {Heymans} C.,  {Jahnke} K.,  {Jogee} S.,  {Meisenheimer} K.,  {Peng} C.~Y.,
  {S{\'a}nchez} S.~F.,  {Somerville} R.~S.,    {Wisotzki} L.,  2005, \apj, 630,
  771

\bibitem[\protect\citeauthoryear{{Zinnecker}, {Keable}, {Dunlop}, {Cannon} \&
  {Griffiths}}{{Zinnecker} et~al.}{1988}]{Zinnecker1988}
{Zinnecker} H.,  {Keable} C.~J.,  {Dunlop} J.~S.,  {Cannon} R.~D.,
  {Griffiths} W.~K.,  1988, in {Grindlay} J.~E.,  {Philip} A.~G.~D.,  eds, The
  Harlow-Shapley Symposium on Globular Cluster Systems in Galaxies Vol.~126 of
  IAU Symposium, {The Nuclei of Nucleated Dwarf Elliptical Galaxies - are they
  Globular Clusters?}.
p.~603

\end{thebibliography}

\end{document}